\documentclass{aa}  
\usepackage{graphicx}
\usepackage{txfonts}
\usepackage{xcolor}
\usepackage{multirow}
\usepackage{float}
\usepackage{scalerel}
\usepackage{tikz}
\usetikzlibrary{svg.path}
\usepackage{booktabs}
\usepackage[normalem]{ulem}

\definecolor{orcidlogocol}{HTML}{A6CE39}
\tikzset{
  orcidlogo/.pic={
    \fill[orcidlogocol] svg{M256,128c0,70.7-57.3,128-128,128C57.3,256,0,198.7,0,128C0,57.3,57.3,0,128,0C198.7,0,256,57.3,256,128z};
    \fill[white] svg{M86.3,186.2H70.9V79.1h15.4v48.4V186.2z}
                 svg{M108.9,79.1h41.6c39.6,0,57,28.3,57,53.6c0,27.5-21.5,53.6-56.8,53.6h-41.8V79.1z M124.3,172.4h24.5c34.9,0,42.9-26.5,42.9-39.7c0-21.5-13.7-39.7-43.7-39.7h-23.7V172.4z}
                 svg{M88.7,56.8c0,5.5-4.5,10.1-10.1,10.1c-5.6,0-10.1-4.6-10.1-10.1c0-5.6,4.5-10.1,10.1-10.1C84.2,46.7,88.7,51.3,88.7,56.8z};
  }
}

\newcommand\orcidicon[1]{\href{https://orcid.org/#1}{\mbox{\scalerel*{
\begin{tikzpicture}[yscale=-1,transform shape]
\pic{orcidlogo};
\end{tikzpicture}
}{|}}}}

\usepackage{hyperref}

\begin{document} 

\title{Astrophotometric search for massive stars in the Milky way. Confronting Random Forest predictions with available
spectroscopy}

\author{N. Monsalves\inst{1,2\orcidicon{0000-0002-4129-8195}}, A. Bayo\inst{2\orcidicon{0000-0001-7868-7031}}, M. Jaque Arancibia\inst{1\orcidicon{0000-0002-8086-5746}}, J. Bodensteiner\inst{3,2\orcidicon{0000-0002-9552-7010}}, A. G. Caneppa\inst{4\orcidicon{0009-0003-3496-2611}}, P. S\'anchez-S\'aez\inst{2\orcidicon{0000-0003-0820-4692}}, R. Angeloni\inst{5\orcidicon{0000-0001-7978-7077}}}

\institute{$^{1}$Departamento de Astronom\'ia, Universidad de La Serena, Avenida Juan Cisternas 1200, La Serena, Chile \\
$^{2}$European Southern Observatory, Karl-Schwarzschild-Str. 2, D-85748 Garching, Germany \\
$^{3}$Anton Pannekoek Institute for Astronomy, University of Amsterdam, Science Park 904, 1098 XH Amsterdam, the Netherlands \\
$^{4}$Instituto de Física y Astronomía, Universidad de Valparaíso, Avenida Gran Bretaña 1111, Valparaíso, Chile \\
$^{5}$Gemini Observatory, NSF's NOIRLab, Av. J. Cisternas 1500 N, 1720236 La Serena, Chile\\
\email{nicolas.monsalves@userena.cl}
}
\date{}


  \abstract
   {Massive stars play a significant role in different branches of astronomy, from shaping the processes of star and planet formation to influencing the evolution and chemical enrichment of the distant universe. Despite their high astrophysical significance, these objects are rare and difficult to detect. With Gaia’s advent, we now possess extensive kinematic and photometric data for a significant portion of the Galaxy that
can unveil, among others, new populations of massive star candidates. In order to produce bonafide bright (G magnitude < 12) massive star candidate lists (threshold set to spectral type B2 or earlier but slight changes in this threshold also explored) in the Milky Way subject to be followed up by future massive spectroscopic surveys, we have developed a Gaia DR3 plus literature data based methodology. We trained a Balanced Random Forest (BRF) with the spectral types from the compilation by \cite{2014yCat....1.2023S} as labels. Our approach yields a completeness of $\sim80\%$ and a purity ranging from $0.51 \pm 0.02$ for probabilities between 0.6 and 0.7, up to $0.85 \pm 0.05$ for the 0.9–1.0 range. To externally validate our methodology, we searched for and analyzed archival spectra of moderate to high probability (p > 0.6) candidates that are not contained in our catalog of labels. Our independent spectral validation confirms the expected performance of the BRF, spectroscopically classifying 300 stars as B3 or earlier (due to observational constraints imposed in the B0-3 range), including 107 new stars. Based on the most conservative yields of our methodology, our candidate list could increase the number of bright massive stars by $\sim$50\%. As a byproduct, we developed an automatic methodology for spectral typing optimized for LAMOST spectra, based on line detection and characterization guiding a decision path.}

\titlerunning{Hunting Massive Stars}
\authorrunning{N. Monsalves et al.}

\maketitle
\nolinenumbers

\section{Introduction}\label{sec_introduction}	

It is commonly agreed that the galactic stellar mass distribution peaks at $\sim0.5 M_{\odot}$, and that its shape has a certain universality \citep{Bastian2010}. However, the slopes towards the higher and lower mass ends are still under debate. Despite their lower numbers in our Galaxy, the effect of massive stars ($M > 8M_{\odot}$) on their surroundings is critical. These stars play a fundamental role in astrophysics, from shaping the processes of star and planet formation \citep{Bally2005} to providing us with key knowledge of the distant universe \citep{Eldridge2022}. In terms of direct outcomes, massive stars can produce a wide range of phenomena, such as Type II supernovae \citep{2009ARA&A..47...63S}, neutron stars and black Holes \citep{Heger_2003}, gamma-ray bursts \citep{Izzard2004,Nakar2007}, and due to their high multiplicity fraction \citep{Sana2012, Moe2017, Barba2017}, they are sources of gravitational waves \citep{Schneider2001}.

Observational classification of stars regarding their mass cannot be directly achieved (as the mass is not an observable). However, classification systems like the MK \citep{1973ARA&A..11...29M} aim to map observables (spectral types) to masses. This system is anchored with a set of standard stars, and classification relies on comparing spectral morphology with these standards. The system focuses on the blue part of the optical spectrum, which historically were the regions of highest sensitivity in photograph plates \citep{1890AnHar..27....1P, Walborn1972,2009ssc..book.....G, sota2011, sota2014}. Regarding massive stars, the 8~$M_{\odot}$ boundary corresponds to spectral type B2 for dwarfs (V), B5 for giants (III), and B9 for supergiants (Ia/Ib, \citealt{morgan51,2009ssc..book.....G,Jesus2024}).

Focusing mostly on the Milky Way, extensive efforts have been made to identify and spectroscopically classify massive stars. These include, among others, targeted studies \citep{sota2011,IACOB2011,2011sca..conf..255S,Goss2013,sota2014,2015hsa8.conf..576S,Li_2021}, specialized compilations such as \citet{2021MNRAS.504.2968P}, and broader spectral type compilations \citep{2014yCat....1.2023S}. Combined, these works yield a total of $\sim$7500 stars spectroscopically classified as B2 or earlier.

This low figure has been subject of a heated debate for several decades. \citet{Kroupa2003} reported an apparent lack of massive stars in the field when compared to their relative number with respect to late-type stars in clusters. This finding is supported by observational studies such as \citet{Rybizki2015}, however, due to the low number and non-homogeneous distribution of high-mass stars, the slope of the mass function above 10~$M_{\odot}$ is an extrapolation. One possible explanation for the apparent deficit of field massive stars is provided by the Integrated Galaxy initial mass function theory \citep{Kroupa2024}, but see \cite{cervino2013a}, and \cite{cervino2013b} for statistical counterarguments. It is therefore necessary to obtain more complete and robust censuses of high-mass stars.

Astronomy is undergoing a transformative phase with an unprecedented influx of data from large-scale surveys. The Gaia mission with its Data Release 3 (Gaia DR3, \citealt{GaiaCollaboration2023}) has delivered six-parameter astrometric data for nearly 900 million stars. This information will soon be complemented by upcoming spectroscopic surveys, for instance, the 4-metre Multi-Object Spectroscopic Telescope (4MOST) \citep{Jong2016,jong2019}, WHT Enhanced Area Velocity Explorer (WEAVE) \citep{weave2024}, Sloan Digital Sky Survey-V (SDSS-V) \citep{SSDSV2019}. Additionally, VLT/MOONS \citep{moons2022} will be equipped with 1000 fibers, significantly more than any existing near-infrared spectrograph. Therefore, to fully and efficiently exploit these extensive capabilities, it is essential to inform the surveys with the most robust candidate lists.

Within this context, \cite{zari2021} presented a comprehensive catalog of hot, luminous stellar candidates with estimated spectral types from O to A in the Milky Way. An immediate goal of the work was to provide robust candidates for SDSS-V follow-up \citep{2023ApJS..266...10K}. The candidate catalog consisted of $\sim$990000 sources with $G < 16$ mag selected with Gaia Early Data Release 3 (Gaia eDR3; \citealt{2021A&A...649A...6G}) photometry and astrometry and Two Micron All Sky Survey (2MASS; \citealt{2006AJ....131.1163S}) photometry. The sample was designed to be more complete than pure with a reported purity of 20\% for stars with temperatures consistent with spectral type B7V.

In this work, we present a supervised machine learning approach designed to improve the yield of massive star candidates with spectral types B2 or earlier in the Milky Way. The method builds on and updated version (fully Gaia DR3) the selection criteria proposed by \cite{zari2021}. In order to validate our approach, we mined spectroscopic archival data for our candidates. As a byproduct, we derived a decision path for the classification of low-resolution spectra of massive star candidates. The path relies on the equivalent width and the presence or absence of specific spectral lines.

This paper is organized as follows: in Section ~\ref{sec:data}, we describe the parental catalog along with its characteristics, and we describe the sources of spectroscopic labels. In Section ~\ref{Sec:ML}, we explain the methodology employed to identify moderate to high probability candidates. In Section \ref{section:results} we present the BRF results and the moderate to high probability candidates of spectral type B2 or earlier. In Section \ref{sec:Discussion}, we discuss the validation of our methodology, including external spectroscopic diagnostics and the comparison with databases and catalogs specialized to massive stars. Finally, in Section~\ref{sec:summary}, we present the conclusions and summarize the main results of this work.
\begin{figure*}
\centering
\includegraphics[scale=0.5]{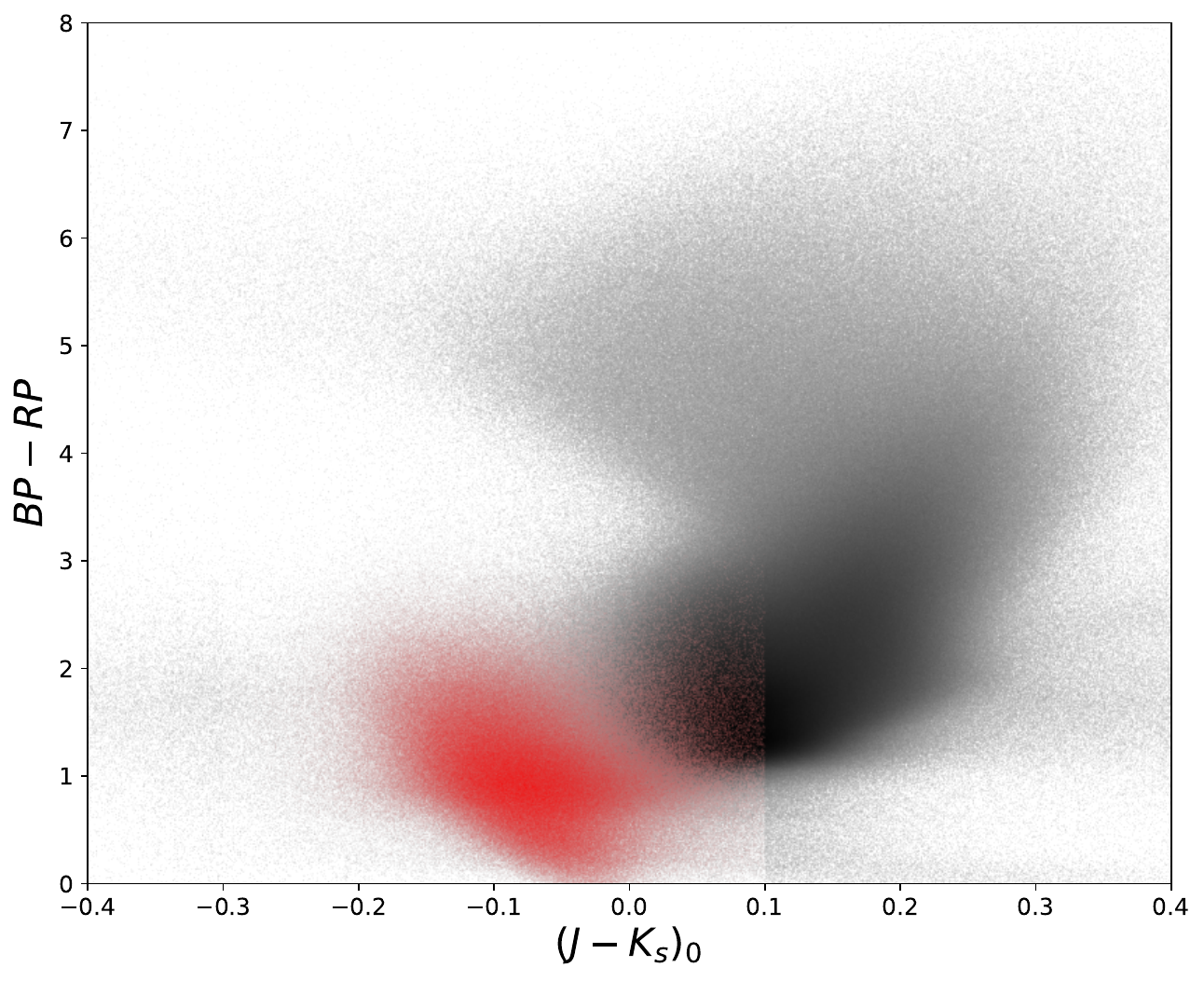}
\caption{Recreation of the color-color diagram of \cite{zari2021} adapted to the full DR3 from Gaia. The entire population brighter than \( G_{\text{mag}} < 16 \) is shown in black, while the OBA GDR3 astrophotometric sample is depicted in red. In this plot, it is easy to distinguish that the brighter stars occupy specific positions in the color-color diagram.
}
\label{fig:zari_sample}
\end{figure*}

\section{The data}\label{sec:data}
In the following we present the different sets of data gathered to develop our classification flow, namely, the source of attributes and lables.

\subsection{OBA GDR3: Parental catalog}\label{sec:OBAsample}

The \cite{zari2021} catalog consisted of sources in Gaia Data Release 2 (Gaia DR2, \citealt{2018A&A...616A...1G} ) crossmatched with 2MASS (see Section 2.1 of \citealt{zari2021}). Posterior to the matching, the authors updated the Gaia photometry and astrometry to those of the Gaia Early Data Release 3 (Gaia eDR3,  \citealt{2021A&A...649A...6G}). Their selection criteria filtered stars brighter than $G < 16$ mag, and then applied simple photometric cuts to select stars of approximate spectral type OBA.
 
To define our parental catalog, we took advantage of the availability of Gaia DR3 and the existing Gaia DR3–2MASS crossmatch\footnote{\href{https://gea.esac.esa.int/archive/documentation/GDR3//Catalogue_consolidation/chap_crossmatch/sec_crossmatch_externalCat/}{ESAC's external crossmatch} catalog service}. We applied their selection function to this updated release. From now on, we will refer to this catalog as OBA GDR3, which contains 1016743 entries, $\sim$3\% more than OBA GR2. This difference maps naturally to the difference in ``seed" catalog as Gaia DR3, as a whole, contains $\sim7\%$ more sources than Gaia DR2. \

Figure \ref{fig:zari_sample} presents a color-color diagram of the entire $G < 16$ mag population in Gaia DR3, together with the OBA GDR3 subsample corresponding to a slightly updated version of Fig. 2 in \cite{zari2021}.

\subsection{OBA GDR3: Features description} \label{sec:features}
We used photometric, astrometric, and color information as features in our analysis. Regarding photometry, the Gaia bands ($G$, $G_{BP}$, and $G_{RP}$)  covered the optical domain and the 2MASS ones ($J$, $H$, and $K_{s}$), the near-infrared. In addition, we considered 15 color indexes resulting from the combination (differences) of pairs of photometric bands. The astrometric parameters considered were parallax ($\overline{\omega}$), proper motion in right ascension ($\mu_{\alpha}$) and declination ($\mu_{\delta}$) with their respective errors ($\sigma_{\overline{\omega}}$, $\sigma_{\mu_{\alpha}}$, $\sigma_{\mu_{\delta}}$).

Finally, we incorporated six quality indicators that measure the goodness of fit of the astrometric solution, such as the \texttt{astrometric\_sigma5d\_max} (ASM), \texttt{astrometric\_excess\_noise} (AEN), Renormalised Unit Weight Error (RUWE), \texttt{astrometric\_gof\_al} (AGOFAL), \texttt{ipd\_gof\_harmonic\_amplitude} (IPDGOFHA), and \texttt{astrometric\_excess\_noise\_sig} (AENS). Extreme values of these features have been interpreted in the literature as sings of multiplicity \citep{2020MNRAS.496.1922B, 2024MNRAS.527.3076D}. For more detailed information on these parameters, refer to the Gaia Data Release 3 documentation.\footnote{\href{https://gea.esac.esa.int/archive/documentation/GDR3/Gaia_archive/chap_datamodel/sec_dm_main_source_catalogue/ssec_dm_gaia_source.html}{GDR3 Documentation}}.

\subsection{OBA GDR3: Spectral type compilation} \label{sec:labels}

The aim of this work is to identify reliable early type candidates based on the features presented in the previous section. We tackle this problem as a supervised classification one with a taxonomy based on spectral types. Obviously, not all objects in the OBA DR3 catalog have spectral type determinations, but a subset do and those will compose the ``universal truth" that we want to propagate with our methodology. The spectral types will be referred to as \textit{labels} from now on, and we used two sources for them: the spectroscopic catalog from \cite{2014yCat....1.2023S}, hereafter referred to as Skiff and wider and very commonly used database, Simbad \citep{2000A&AS..143....9W}. We note that Simbad is a manually curated survey that may include mismatches and fall behind the current literature; however, it is the most complete stellar database available at present.

To assign labels to the entries of the OBA GDR3 catalog, we performed two independent crossmatches: to Simbad and Skiff. In both cases, we keep every entry from OBA GDR3 and employed a 2\arcsec radius for coordinate matching.

Regarding multiple counterparts, over 95\% of targets in Simbad had a one to one correspondence in the defined radius, and hence, in case of multiple-match, we opted to use as ``true" counterpart, the closest target in terms of angular distance. The situation was drastically different in the matching with Skiff, where over 70\% of the targets have indeed multiple counterparts. This was expected, as Skiff, by construction, does not prevent multiple entries for the same astrophysical object, in Sec.~\ref{sec:RepEntries} we present our approach to deal with duplications.

\subsubsection{Numerical encoding}\label{sec:cod_spec_types}

The labels provided in Skiff,  and Simbad are alphanumerical quantities that we proceeded to encode in numerical tuples to preserve the ordered relation of the labels (an O stars is closer in properties to a B star than to an A and so forth). These encodings are used in Section \ref{Sec:ML} to train the model and revised in Section \ref{Sec:Discussions} to assess possible limitations.

For the first element of the tuple, the letter, O to M, was mapped to integer values in increments of 10, ranging from 0 to 60. If a subtype was reported in either catalog, its integer value (0–9) was added. Otherwise, in cases where a subtype was not present, an integer was sampled from a uniform distribution of possible values and then added. For instance, a star classified only as type B was initially encoded as 10. We then sampled a random integer from 0 to 9; if the result was 1, the final encoded label was set to 11, corresponding to B1. We noticed that this may introduce noise in the labels; however, this uncertainty with the subspectral types only affects 1.6\% of the labels.

In addition, whenever uncertain values or ranges (e.g., B9/A0, F1-F3) were reported, the individual encoding for the extremes were computed and used as the inclusive limits for a distribution to be uniformly sampled. It must be noted that this sampling artificially ``produces" spectral types for which no spectra of standard stars may exist \citep{Jesus2024}, however, the classifier (presented in Sec.~\ref{Sec:ML}) will be trained on a lightened binary taxonomy ``earlier or later than B2", and hence this (ease of computation motivated) artifact can be neglected. Finally, when the uncertainty in the spectral classification spanned more than one main spectral type (e.g., OB), the label was saved as a known label but excluded from the training, while targets with reported composite spectral types (such as kA2hA9mF2\footnote{This corresponds to A2 for the Ca K-line subtype (k), A9 for the hydrogen-line type (h), and F2 for the metallic-line type (m)}
 ) were discarded.

Although Table \ref{table:peculiarities} provides the full encoding of the spectral types here described, when training the Random Forest classifier we simplified our labels to two: 0 for encoded spectral type > 12 (`negative" class), and 1 for encoded spectral type $\le$ 12 (`positive" class).

Since the different elements used as attributes among themselves (photometry and kinematics) and labels are not coetaneous, for objects exhibiting extreme variability this could pose some source of noise in the modeling. However, we expect the number of such cases to be negligible. For example, the fraction of extreme variability cases reported in the VISTA Variables in the Via Lactea (VVV) survey \citep{2010NewA...15..433M} is below $10^{-5}\%$ \citep{2024MNRAS.528.1789L}.
\subsubsection{Addressing repeated entries}\label{sec:RepEntries}

While the (2\arcsec radius) crossmatch of OBA GDR3 data with Skiff may yield multiple entries for the same astronomical sources, either due to different sources or to multiple spectral classifications of the same source, the crossmatch with SIMBAD does not. Thus, in order to faithfully compare both crossmatched datasets against each other in the following subsection, we first performed an internal match within the Skiff-correlated set with respect to its astrometric coordinates, as to identify and group multiple records that describe the same object. The match threshold criterion was a maximum radius of 2\arcsec of angular distance. For each group, we computed the mean and standard deviation of their spectral type encodings. Specifically, grouped entries with standard deviation values greater than $1.0$, $3.0$, and $5.0$ (i.e., $1$, $3$, and $5$ units of spectral subtype) were discarded, resulting in three filtered sets. This standard deviation will hereafter be referred to as $\sigma_{\text{literature}}$.

\begin{figure}
\centering
\includegraphics[scale=0.7]{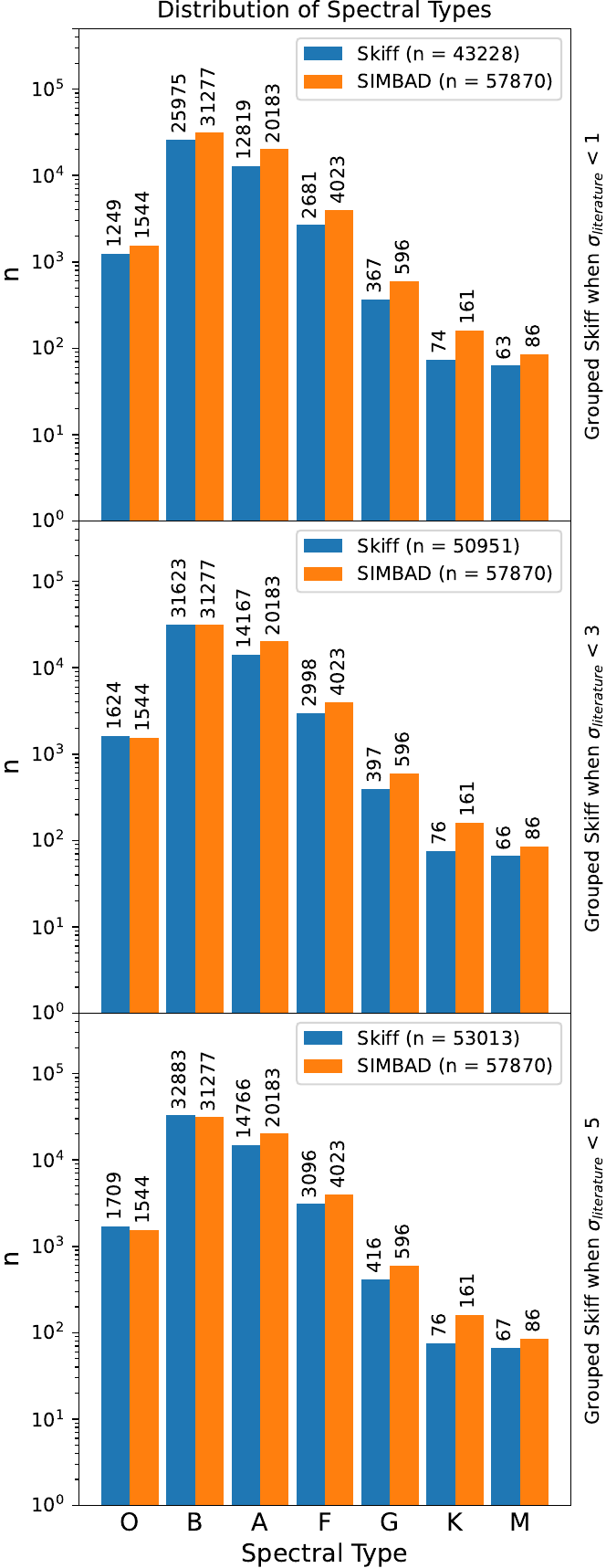}
\caption{Distribution of OBAFGKM spectral types for Skiff and Simbad. The subplots from top to bottom correspond to different $\sigma_{\text{literature}}$ thresholds for Skiff (see text).}
\label{fig:comparation_simbad_skiff_spdist}
\end{figure}

\subsubsection{Comparing the two sources of labels}

We compared Simbad and Skiff in detail to assess the differences between these two databases. Fig. \ref{fig:comparation_simbad_skiff_spdist} shows the number of spectral types recorded in Skiff and Simbad, considering the previously discussed $\sigma_{\text{literature}}$. We collected spectral type information for approximately 55000 stars, which represented $5\%$ of the OBA GDR3 sample. In all cases, Simbad contains more stars with spectral type information than Skiff. For O stars in Skiff, the number increases by $30\%$ when the $\sigma_{\text{literature}}$ threshold rises from 1 to 3 and by $5\%$ when it increases from 3 to 5, with a similar trend observed for other spectral types. Based on this, we adopt a $3-\sigma_{\text{literature}}$ threshold for grouping Skiff data, achieving a balance between robustness and sample size.

The complete description of the encoding for the luminosity class can be found in Appendix \ref{apendixcomplement}. On Fig. \ref{fig:comparation_simbad_skiff_luminosity}, however, we do incorporate this information in the comparison between Simbad and Skiff to offer a complete view. As can be seen in the figure, both databases exhibit a similar distribution, with dwarfs being the most abundant class. However, there is a discrepancy in the number of subgiants (class IV), due to uncertainties in the luminosity classifications in Skiff (e.g., IV/V or V/IV), which are encoded as intermediate between these classes (see Section \ref{sec:cod_spec_types}). When combined with the binning process used in the histogram, these uncertainties result in a higher contribution for subgiants.

\begin{figure}[htpb]
\centering
\includegraphics[scale=0.66]{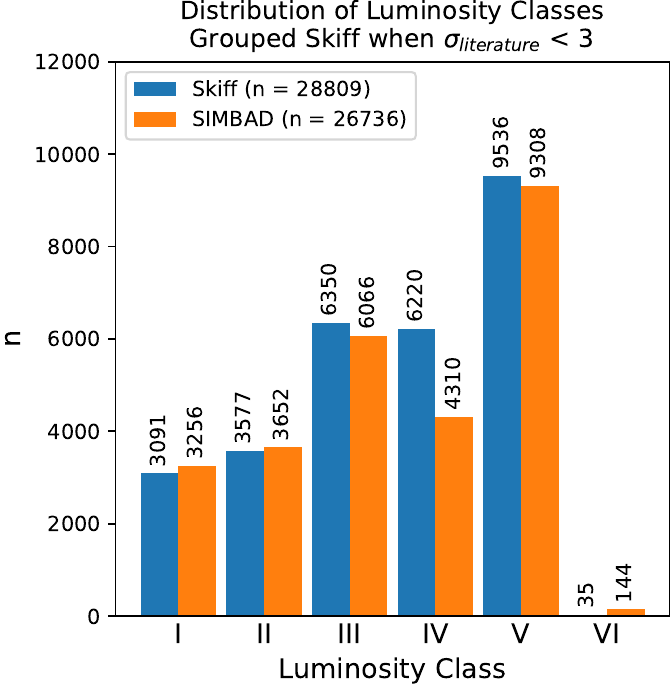}
\caption{Distribution of luminosity classes for Skiff and Simbad. The luminosity class reported for Skiff is based on a $\sigma_{\text{literature}}$ threshold of 3.}
\label{fig:comparation_simbad_skiff_luminosity}
\end{figure}

\begin{figure}[htpb]
\centering
\includegraphics[scale=0.6]{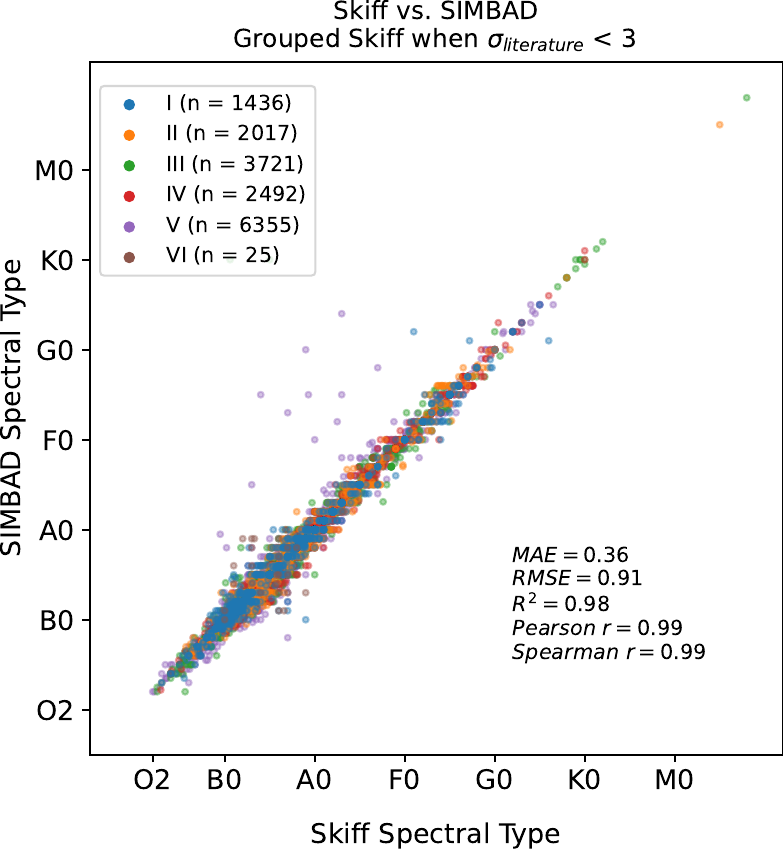}
\caption{Comparison of spectral types for Skiff and Simbad. The different colors represent luminosity classes. Only luminosity classes that coincide in both databases were considered.}
\label{fig:Skiff_vs_Simbad_sd_3_with_pecs}
\end{figure}

Figure \ref{fig:Skiff_vs_Simbad_sd_3_with_pecs} compares the labels identified for the same object in the two databases for those cases where the luminosity class is in agreement. We included various dispersion metrics for comparison: the Mean Absolute Error (MAE), Root Mean Square Error (RMSE), Coefficient of Determination ($R^2$), Pearson Correlation Coefficient ($r$), and Spearman's Rank Correlation Coefficient ($\rho$), available in \cite{scikit-learn}. The MAE and RMSE were 0.36 and 0.91, respectively, with differences remaining below one spectral subtype unit. The other metrics were very close to 1, confirming a strong correlation between the two databases and indicating that they were highly comparable.

We identified a few outliers above the bulk of the data and investigated them using both the SIMBAD and Skiff databases. In most cases, we could not retrieve the reference listed in SIMBAD. In a few others, we found misreferences in Skiff or more updated spectral classifications in one of the two databases. We did not exclude them from our analysis, since all these stars are later than B2 and represent only a very small fraction of the labeled data.

\begin{figure}
\centering
\includegraphics[scale=0.6]{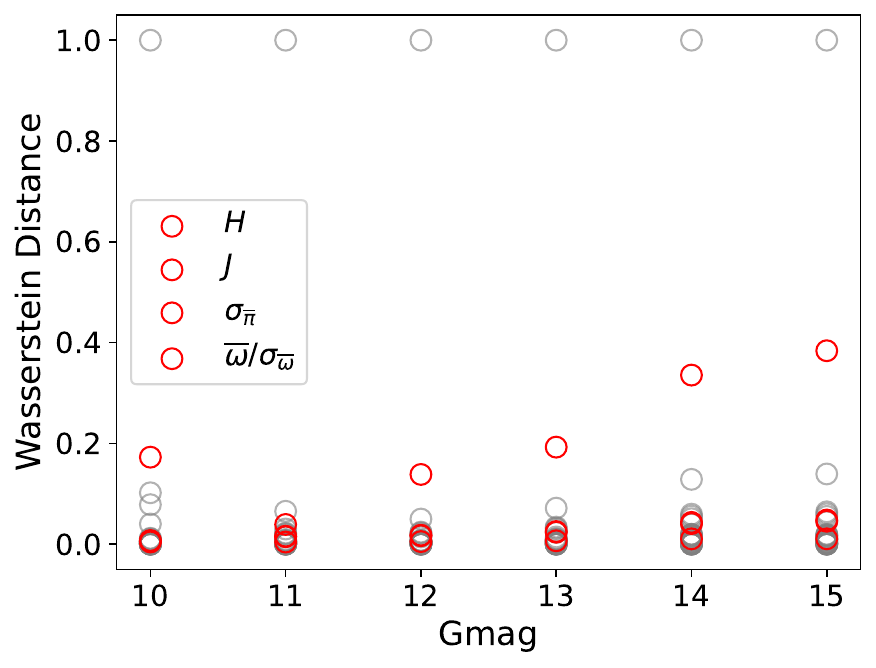}
\caption{Normalized Wasserstein distance between the OBA encoded labels and the pool of new candidates for each feature per $G$ band magnitude. Each feature is shown in grey, while the four most important features for the model (see Section~\ref{section:results}) are highlighted in red.}
\label{fig:WassertsteinDistance}
\end{figure}

\begin{figure}
\centering
\includegraphics[scale=0.45]{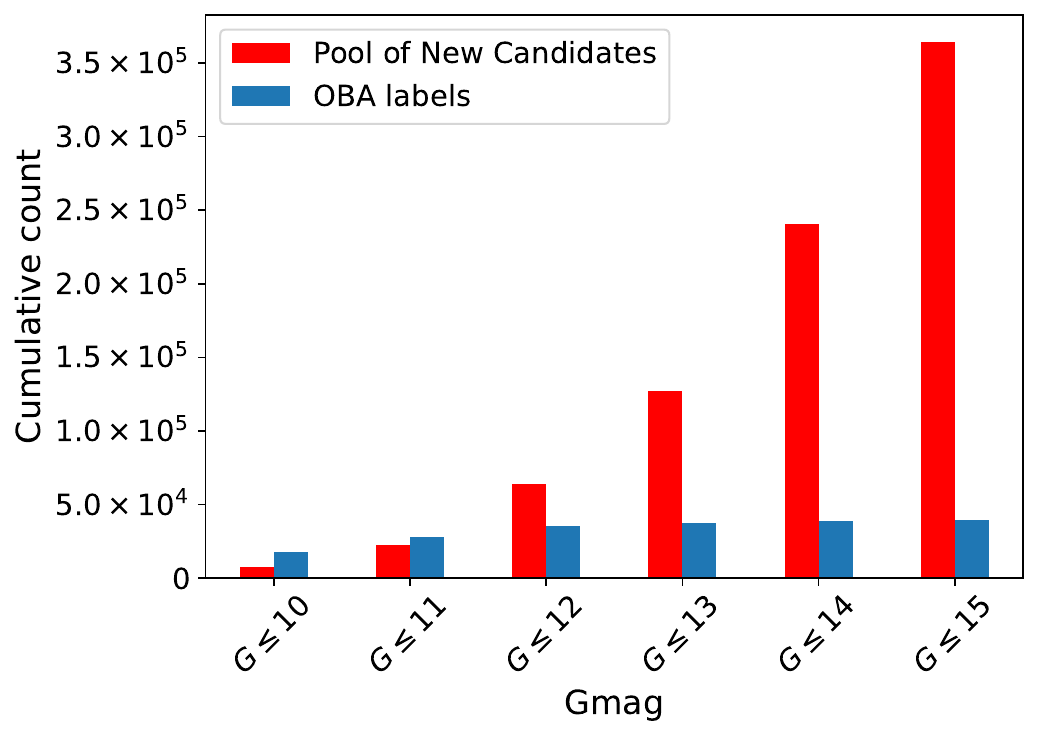}
\caption{Cumulative number of sources for different $G$ band magnitude cuts for the OBA encoded labels and the pool of new candidates.}
\label{fig:CumulativeCounts}
\end{figure}

Given this strong correlation, we ultimately decided to use the Skiff database for our analysis. This choice is motivated because approximately $\sim$1000 objects in Simbad lack references for their spectral types; $\sim$7000 objects potentially suffer from suboptimal astrometry, with a positional precision of $8$–$12\arcsec$ compared to the $\sim1\arcsec$ coordinate precision guarantee in Skiff. Finally, Skiff ensures spectroscopic origins for its compilation, whereas Simbad does not. Although it was possible to combine classifications from both databases to increase the sample size, this approach could have introduced additional dispersion or variance due to the inclusion of objects classified through different techniques.

\subsubsection{Simplification of encoded labels and adopted threshold}\label{sec:simplabel}

We simplified the encoded labels into more global and representative categories. Since the fundamental truth sample is meant to provide 'standard examples' of each class, we filtered out phase-dependent spectral types (e.g., OBe), stellar spectra with reported uncertainties in the classification, and non-MK classifications (e.g., Wolf-Rayet stars, white dwarfs, planetary nebulae, and carbon stars). Additionally, 'em' designations were removed, as they lack information about the primary spectral type. Lastly, we excluded the $\sim$2000 systems reported in the catalog as multiples, due to the limited number of cases (only 2\%) in which spectral types for both components are provided. This decision could be seen as at odds with the fact that multiplicity is known to scale with primary mass. However, the uncertainty introduced by the lack of individual spectral types (and hence across the period - mass-ratio parameter space) for these systems was deemed detrimental to the training process. 
 
The "boundary" between massive and not massive stars is not a simple cut in terms of spectral type for B stars, as the luminosity class has to be considered. However, more than half of the available encoded labels lack this information. Promoting purity over completeness, we set B2 as the threshold, as this spectral type corresponds to a massive star across all luminosity classes \citep{Jesus2024}.

\begin{table}[htpb]
  \centering
  \caption{Number of classifications excluded}
  \label{table:simplification}
  \begin{tabular}{lc}
    \toprule
      Spectral type (labels) & Number \\
    \midrule
    Binaries & 2456 \\
    em & 6220	 \\
    OB & 16363 \\
    OBe & 721 \\
    Non MK and uncertainty & 16345 \\
    \bottomrule
  \end{tabular}
  \tablefoot{These are not necessarily unique stars, as the same star can appear in different works, each reporting a (potentially) different spectral type}
\end{table}

 \section{Methodology: Balanced Random Forest} \label{Sec:ML}
In this section, we used the simplification of the encoded labels introduced in section \ref{sec:labels} and the features detailed in section \ref{sec:features} to address the classification of massive stars as a supervised learning problem. We applied a Balanced Random Forest (BRF), suitable for handling imbalanced training datasets \citep{chen2004using}. We trained the \textit{BalancedRandomForestClassifier} method from the \textit{imbalanced-learn}\footnote{\href{https://imbalanced-learn.org/stable/references/generated/imblearn.ensemble.BalancedRandomForestClassifier.html}{BalancedRandomForestClassifier}} using the simplification of the encoded labels from Skiff within the OBA GDR3 sample and applied the classification to the unlabeled data.

 \subsection{Preprocessing Features}

We removed sources with NaN values in any feature or duplicate $J$, $H$, and $K_{s}$ photometry. The percentage of duplicate sources using the \textit{tmass\_oid} identifier was $0.013\%$. Then, we defined the minimum and maximum values for each feature by calculating the 1st and 99th percentiles of the encoded labeled data. We only selected the pool of new candidates that fell within these boundaries. This ensured that all unlabeled candidates were within the specified feature restrictions, as shown in Table \ref{table:min_max_features}.

We aimed to quantify how representative the OBA encoded labeled data are with respect to the pool of new candidates for each feature. In Fig. \ref{fig:WassertsteinDistance}, we present a similarity test using the Wasserstein distance\footnote{\url{https://docs.scipy.org/doc/scipy/reference/generated/scipy.stats.wasserstein_distance.html}} between these samples. In Fig. \ref{fig:CumulativeCounts}, we show the cumulative number of sources per bin of G magnitude.

The similarity test was normalized between the maximum and minimum distance for each bin of magnitude. The most distinct feature across all magnitude bins was \texttt{AENS}, followed by $\sigma_{\overline{\omega}}$, the second most important feature for the model (see Section~\ref{section:results}). This feature reached a minimum at $G = 11$ mag and began to increase for both bright and faint stars. From Fig \ref{fig:CumulativeCounts}, we observed that for bright stars ($G = 10$ mag), the number of OBA encoded labeled sources at $G = 10$ mag is limited, which may explain the larger distance in this magnitude bin. For fainter stars, starting around $G \approx 12$, there was no noticeable increase in the cumulative number of OBA encoded labels. This suggested a bias in the labeled sample toward brighter stars, making extrapolation to fainter magnitudes uncertain. To be more conservative, we selected stars brighter than $G_{\text{mag}} \leq 12$ as a compromise between completeness and purity. The final sample consisted of 99388 stars up to $G = 12$~mag, of which 35465 had spectral types and 63923 were unlabeled.

\subsection{Metrics}

We evaluated the model's performance using Recall, Precision, F1-score, Geometric Mean, and Index Balanced Accuracy (IBA). In the text, we refer to completeness and purity as recall and precision, respectively. The last two metrics, designed for imbalanced datasets, were taken from the \textit{imbalanced-learn} package \citep{JMLR:v18:16-365} \footnote{\url{https://imbalanced-learn.org/stable/references/metrics.html}}. The IBA incorporates any metric ($M$) and allows emphasis on specific classes through a weighting parameter ($\alpha$). These metrics helped assess model construction, final classification, and comparison with other models. The metrics are presented as follows:

\begin{subequations}
\begin{equation}
    \text{Recall} = \frac{TP}{TP+FN} 
\end{equation}
\begin{equation}
    \text{Precision} = \frac{TP}{TP+FP} 
\end{equation}
\begin{equation}
\text{F1 Score} = 2 \times \frac{\text{Precision} \times \text{Recall}}{\text{Precision} + \text{Recall}}
\end{equation}
\begin{equation}
    TNR = \frac{TN}{TN+FP} 
\end{equation}
\begin{equation}
    Dom = recall -TNR
\end{equation}
\begin{equation}
    G\text{-}mean = \sqrt{\text{Recall} \times TNR}
\end{equation}
\begin{equation}
IBA_{\alpha}(M) = (1+\alpha \cdot Dom) \cdot M
\label{iba}
\end{equation}
\label{subecuation}
\end{subequations}

Where TP are the number of true positives, FP the false positives, FN the false negatives and TN true negatives. See subsections~\ref{subsec:buildingmodel} and~\ref{subsec:PerformanceBRF} for the use and results of these metrics. \\

\subsection{Building the model} \label{subsec:buildingmodel}

The random forest (RF) is a tree-based algorithm that consists of an ensemble of decision trees \citep{2001MachL..45....5B}. Each tree is constructed using a random subsample with replacement (a bootstrap sample) from the training set, and the final prediction is determined by a majority vote among all the trees.

The structure of a decision tree consists of a root node, internal nodes, and terminal (leaf) nodes. Data is classified based on decision rules inferred from features at different levels, from the root node (level 0) to the terminal nodes (maximum depth). A terminal node is reached when all training examples within it belong to the same class. These components are defined by hyperparameters: \textit{n\_estimators} indicates the number of decision trees, \textit{max\_features} denotes the subset of features used to determine the best split at each node, and \textit{max\_depth} sets the limit for recursive division.

Traditional RFs show high performance in different problems, however they underperform on highly imbalanced datasets since bootstrap samples often underrepresent the minority class. Instead, we used the Balanced Random Forest (BRF), a variation of RFs that modifies the sample distribution to ensure equal class representation in each tree. We selected the BRF from the \textit{Imbalanced-learn} module, which offers various under-sampling strategies for bootstrapping: \textit{majority}, \textit{not minority}, \textit{not majority}, \textit{all}, and \textit{auto}. We tested independent strategies for binary classification: 'majority', 'not majority', and 'all'.

\begin{table}
    \centering
    \caption{hyperparameters tested for the Random Forest and the best value}
    \label{table:hyperparametersTested}
    \small
    \begin{tabular}{|l|l|l|}
    \hline
    Features & Tested parameters& Best parameter \\
    \hline
    n\_estimators & [500, 800, 1200] & 1200 \\
    max\_features & [sqrt, log2] & log2 \\
    max\_depth & [10, 20, 30, None] & 30 \\
    sampling\_strategy & [majority, not majority, all] & majority \\
    \hline
    \end{tabular}
\end{table}

To optimize BRF construction, we used Grid Search CV with stratified k-fold cross-validation. This method systematically explores hyperparameter combinations from a predefined grid. It divides the data into 'k' folds, using 'k-1' folds for training and the remaining fold for testing. We used a 5-fold stratified cross-validation and selected the IBA combined with \( Gmean^2 \) as the scoring metric, setting \( \alpha = 0.1 \), as recommended in \cite{GARCIA201213}. Table \ref{table:hyperparametersTested} presents the tested and optimal hyperparameters.

\begin{figure}[htbp!]
\centering
\includegraphics[scale=0.44]{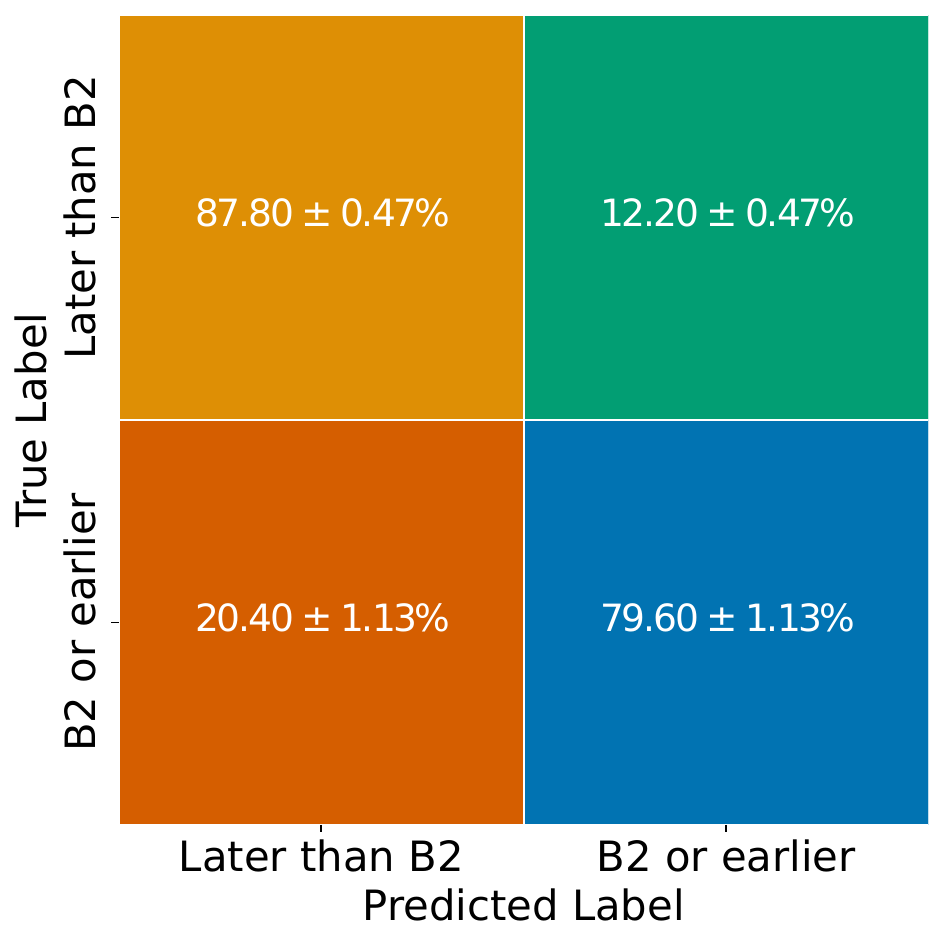}
\caption{Confusion matrix for stars earlier than B2 and later than B2. The confusion matrix was obtained by generating 20 distinct training and testing sets, and training independent models on each. After training, the models were applied to their respective testing sets. We report the mean and standard deviation across all testing sets.}
\label{fig:CM_colored}
\end{figure}

\begin{figure*}[htbp!]
\centering
\includegraphics[scale=0.41]{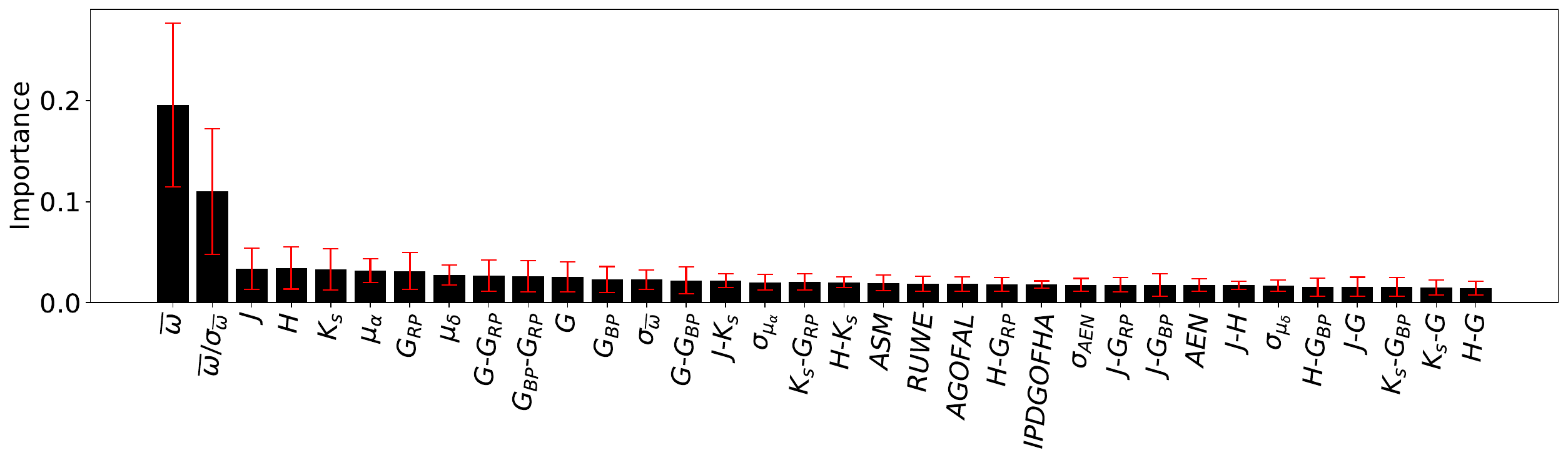}
\caption{Feature importance for the BRF, sorted by significance. The importance is calculated as the mean and standard deviation of the accumulated reduction in variance across all trees and individual splitting iterations.}
\label{fig:features_importance}
\end{figure*}

\section{Results: Candidates from BRF}\label{section:results}

In the following, we use the built model and the preprocessed features to train a BRF and extend the classification to the unlabeled data. We also present the performance metrics of the algorithm and the moderate to high probability massive star candidates.

\subsection{Performance of the BRF} \label{subsec:PerformanceBRF}

We defined the BRF using the best combination of hyperparameters and evaluated the performance of the model roughly following \cite{2021AJ....161..141S}. We created 20 random permutations of the training and test sets using \textit{StratifiedShuffleSplit} from \textit{scikit-learn}, allocating \(80\%\) of the data for training and \(20\%\) for testing in each permutation. This corresponded to 28372 sources in the training set and 7093 in the test set per split. Across the full dataset, 4393 sources were classified as B2 or earlier, and 31072 as later than B2. Subsequently, we trained 20 BRFs, each on a separate splitting iteration using the training sample, and evaluated their performance with the respective test sets. We used a probability threshold of 0.6, above which stars are classified as spectral type B2 or earlier. We did not include oversampling techniques, as the ratio between the majority and minority classes is approximately seven, significantly lower than the $\sim$42 reported in the original SMOTE paper \citep{chawla2002smote}.

\begin{figure*}
\centering
\includegraphics[scale=0.7]{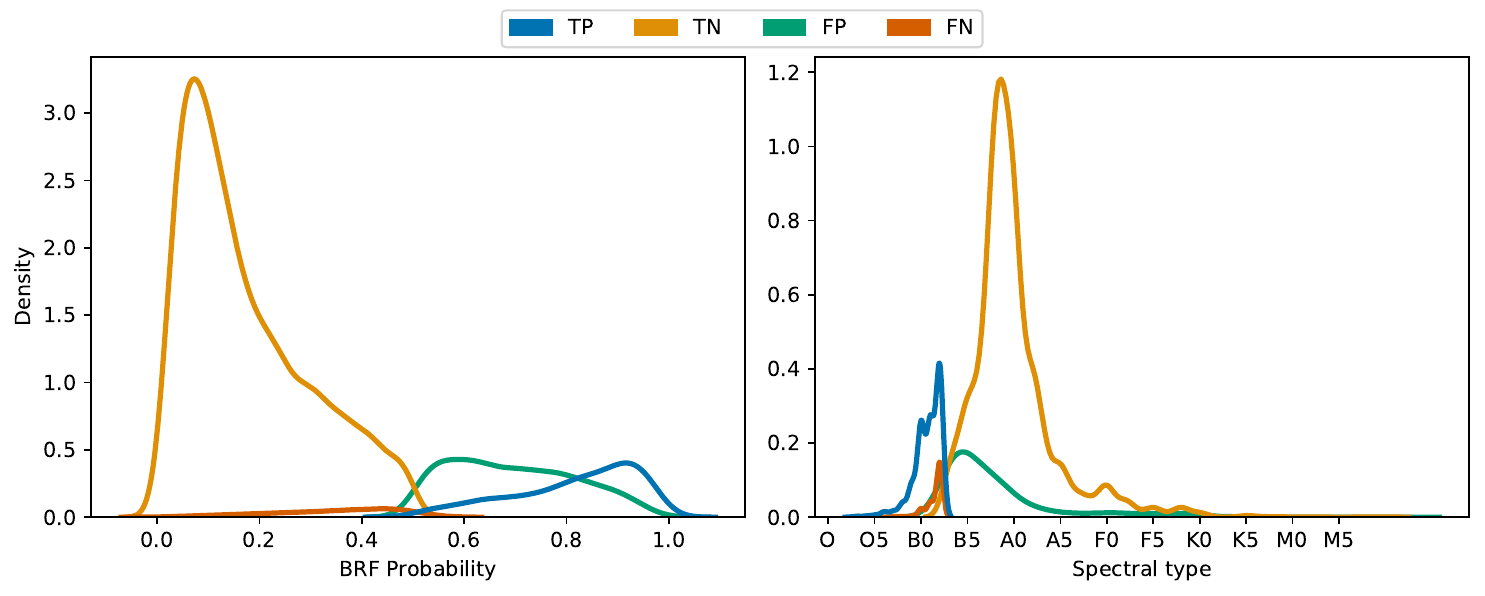}
\caption{Kernel density plots for the metrics in the test set. The left panel shows the distribution in the probability of the BRF. The right panel shows the distribution of the different spectral types}
\label{fig:metric}
\end{figure*}

Figure \ref{fig:CM_colored}  summarizes the results obtained for the 20 random permutations for each test set. TP, TN, FP, and FN values are averaged over these permutations, with the error defined as their standard deviation. We obtained a TN fraction of $\sim88\%$ and an FP fraction of $\sim12\%$. The TP fraction was $\sim80\%$, while the FN fraction was $\sim20\%$. We measured a standard deviation of $0.47\%$ for stars with spectral type later than B2, lower than that of stars with spectral type B2 or earlier.

Figure \ref{fig:features_importance} shows the variations of the relevance of each feature in the deciding power of the 20 models. When the randomness of the different models is taken into account, the relevance of most features are indistinguishable, with two exceptions: the parallax and the parallax-over-error.

The H–G color was the least informative for the model's performance, although it traces the slope of the stellar spectrum. We explored this effect using the spectral energy distributions (SEDs) of 100 nearby sources ($<$1kpc) with high predicted probabilities of being of spectral type B2 or earlier. These SEDs were fitted using \texttt{VOSA} \citep{Bayo2008}, allowing extinction to vary freely within the catalog provided limits. The results showed that H–G is highly degenerate with extinction, which explains its low relevance with respect to effective temperature.

\begin{table}[htpb]
  \centering
  \caption{Performance metrics for the test sets}
  \label{table:metrics}
  \resizebox{0.5\textwidth}{!}{  
    \begin{tabular}{lccc}
      \toprule
         & Precision & Recall & F1 Score  \\
      \midrule
        Later than B2 & 0.968 $\pm$ 0.002 & 0.878 $\pm$ 0.005 & 0.921 $\pm$ 0.003 \\
        B2 or earlier & 0.48 $\pm$ 0.01 & 0.80 $\pm$ 0.01 & 0.599 $\pm$ 0.009 \\
      \bottomrule
    \end{tabular}
  }
\end{table}

We trained the final BRF classifier using a single data split, motivated by the small variation observed in the confusion matrices across different $k$-folds. All features were used to train the model, which was then applied to classify the OBA candidates in GDR3. We used $80\%$ of the data for training and $20\%$ for testing, applying the best hyperparameters from Table \ref{table:hyperparametersTested}. Fig. \ref{fig:metric} shows the kernel density estimates (KDE) for our test sets by metric. The left panel presents the BRF probability, while the right panel displays its distribution across spectral types. The TN peak was near 0.15, transitioning smoothly from lower probabilities before distorting as it approached 0.5, where the model had the highest classification confusion. For stars with probabilities above 0.6, TP and FP clearly overlapped. In the right panel, the metrics were associated with the spectral type, with the TN peak corresponding to A0 stars, while the FP peak was centered around B5 stars.

The peak in FP around spectral type B5 might be explained by the subtleties in spectral classification. The main spectroscopic difference between a B5 and a B2 star is the disappearance of silicon lines (very weak absorption features), accompanied by a strengthening of the Mg II 4481 line. Given that the model is informed only with photometry (“even worse”, broad band photometry), the model has not been exposed to data that traces such subtle differences and, therefore struggles to distinguish between these spectral types. 
 

\begin{figure}
\centering
\includegraphics[scale=0.35]{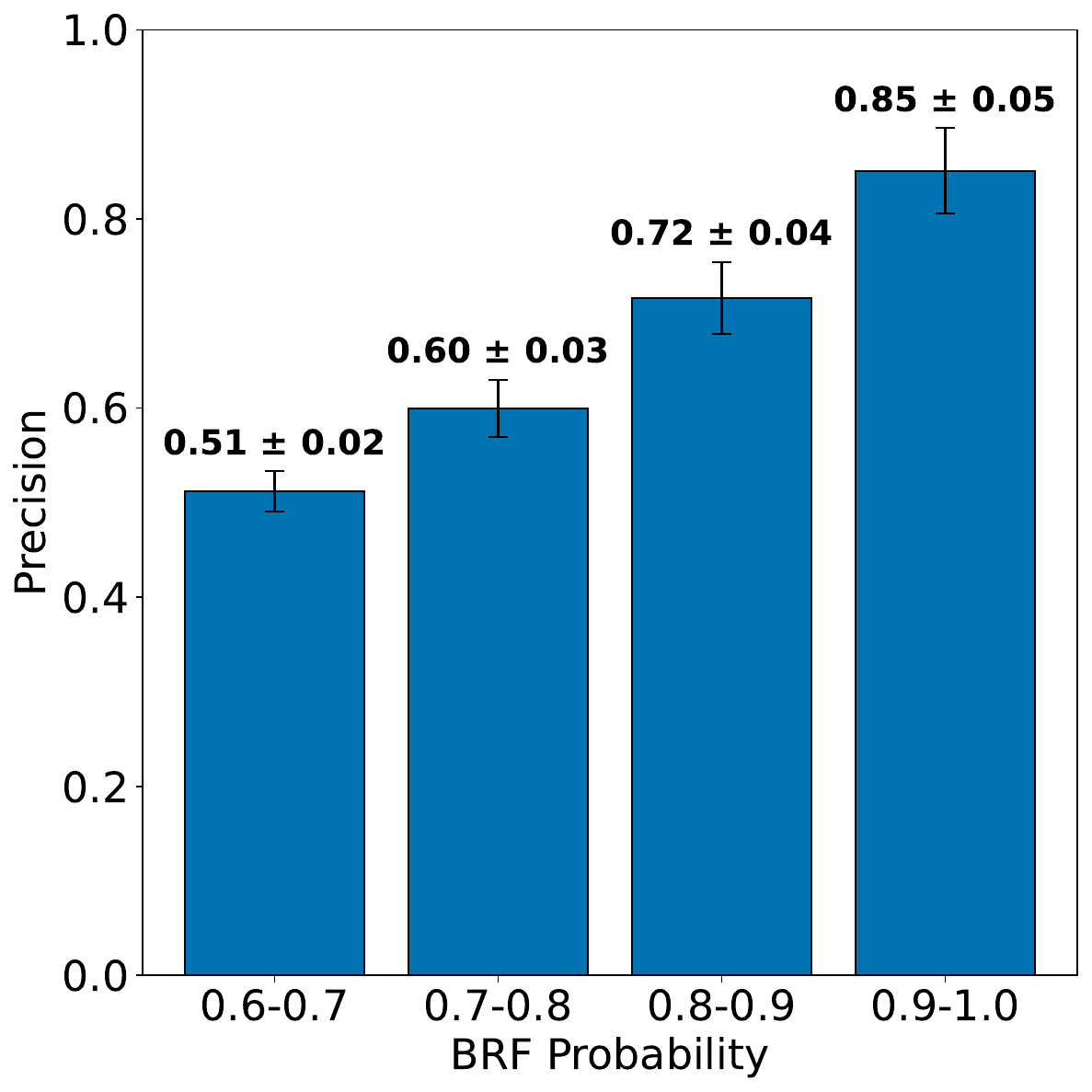}
\caption{BRF precision for the test set in detecting B2 or earlier stars at different probability thresholds}
\label{fig:Precision_dif_thresholds}
\end{figure}

Table \ref{table:metrics} presents the performance metrics for the test sets. We observe high recall for stars with spectral types later than B2 and B2 or earlier. While precision remains high for stars later than B2, it is lower for B2 or earlier stars, indicating a high fraction of false positives. Fig. \ref{fig:Precision_dif_thresholds} shows how precision improves at higher probability thresholds. In addition, as noted in Section \ref{sec:labels}, the classification of massive stars based on spectral type is a working approximation, typically set at B2 or earlier. However, some authors, such as \cite{Zinnecker2007} (see Table 1), have considered B3 or early stars as massive. Therefore, we also presented the results using this alternative threshold in Appendix \ref{apendixB}. Finally, this trade-off between high completeness and lower purity can be addressed, as the final classification of massive stars requires spectroscopic confirmation.

\begin{figure}
\centering
\includegraphics[scale=0.6]{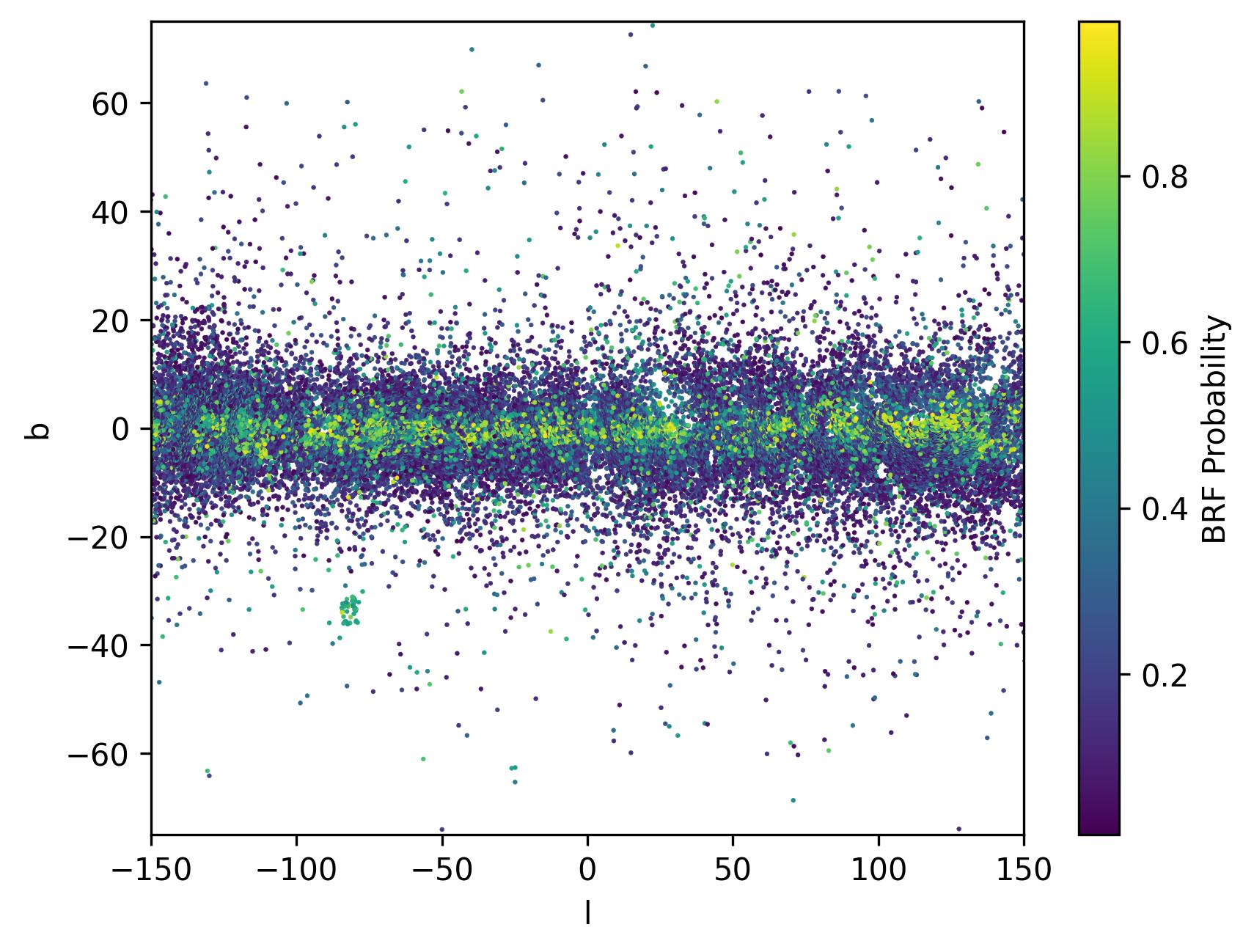}
\caption{Galactic distribution of the candidate samples. This shows the distribution of stars outside our encoded labeled sample. The color represents the probability of being an star with spectral type B2 or earlier.}
\label{fig:galactic_plane}
\end{figure}

\subsection{Moderate to high probability massive stars candidates} \label{subsec:Highprobcandidates}

From the sample of 99 388 stars with $G\leq 12$ mag, we classified the 63 923 unlabeled candidates using the BRF. We show the galactic distribution of our candidates in Fig. \ref{fig:galactic_plane}. The colors indicate the probability of classification as a star with spectral type B2 or earlier. We observed that the highest probabilities are more concentrated near the Galactic plane, even without RA and Dec coordinates as input for the algorithm.

\begin{figure}
\centering
\includegraphics[scale=0.6]{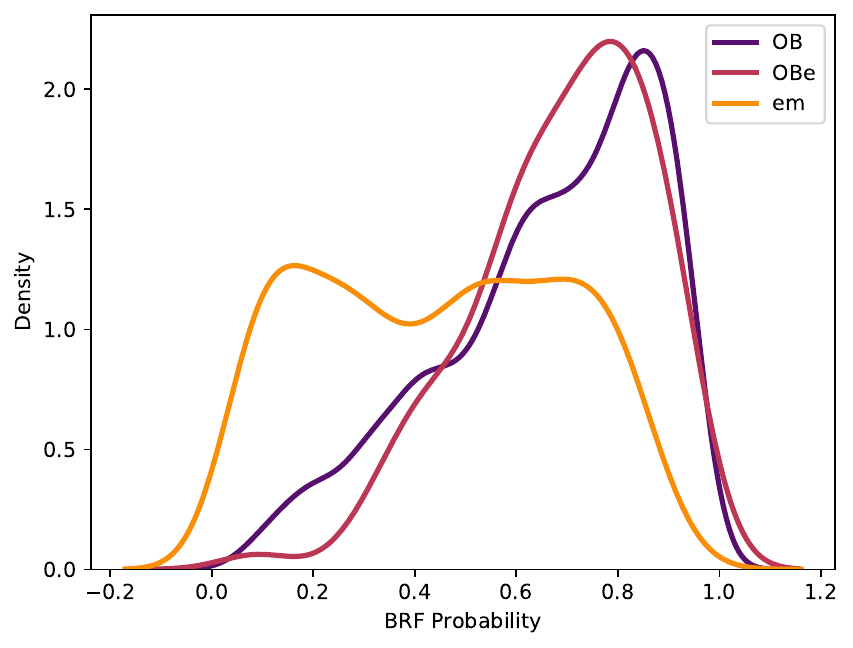}
\caption{Kernel density plots for labeled data that are not included in our label codification.}
\label{fig:noincludes}
\end{figure}

Figure \ref{fig:noincludes} shows the probability distribution for Em, OB, and OBe stars, which are the most abundant simplified encoded labels that were not included in the training or test sets. The distributions for OB and OBe stars are skewed toward higher probabilities, reflecting their predominance as early-type stars. In contrast, the Em distribution shows two peaks at low and high probabilities, suggesting the presence of both early- and late-type stars with emission. These trends suggest that the simplification of the encoded labels captured the global phenomenon without an evident systematic pattern for phase-variable spectral types.

Other spectral classes not considered in the training were excluded due to the feature space cuts applied during sample construction. As a result, we do not have a statistically significant number of objects from these classes to evaluate model performance reliably. In addition, assessing the classifier’s performance on Be stars is dificult because the Gaia and 2MASS photometry, the labels, and the spectroscopic data are non simultaneous. Since emission lines are time dependent features, this prevents us from drawing reliable conclusions. However, we expect Be stars to show distinct patterns in the Gaia feature space \citep{2025MNRAS.539.1964R}.

From the $\sim$ 100000 stars analyzed, we identified 9089 moderate to high probability massive star candidates (probability > 0.6) that were not included in our encoded labeled dataset. According to Table \ref{table:metrics}, this threshold corresponds to an expected precision of $\sim 0.5$ for identifying massive stars. Based on the FP and TP rates shown in Fig. \ref{fig:metric}, we expect the FP to be centered around B5 stars. This sample is more than twice the number of massive stars used for the development of the model, significantly expanding the set of potential massive stars and representing a valuable target for spectroscopic follow-up.

\section{Discussion} \label{sec:Discussion}

In the following we discuss the validation of our methodology incorporating external spectroscopic diagnostics, and the characteristics of our sample with respect to the literature.

\subsection{Multiplicity considerations}

The process of simplification of the numerical encoding of labels discussed in Section~\ref{sec:simplabel}, implied the removal of multiple systems. However, as previously mentioned, multiplicity is a very common feature of massive stars \citep{Offner2023}. In order to assess the performance of our (singled source trained) algorithm in detecting multiple systems with a massive primary, we analysed the predictions of the model for a sample of 29 confirmed binaries from the Skiff catalogue. The number of systems was limited, as we only considered multiple systems that were not used in the previous analysis.

\begin{figure}
\centering
\includegraphics[scale=0.75]{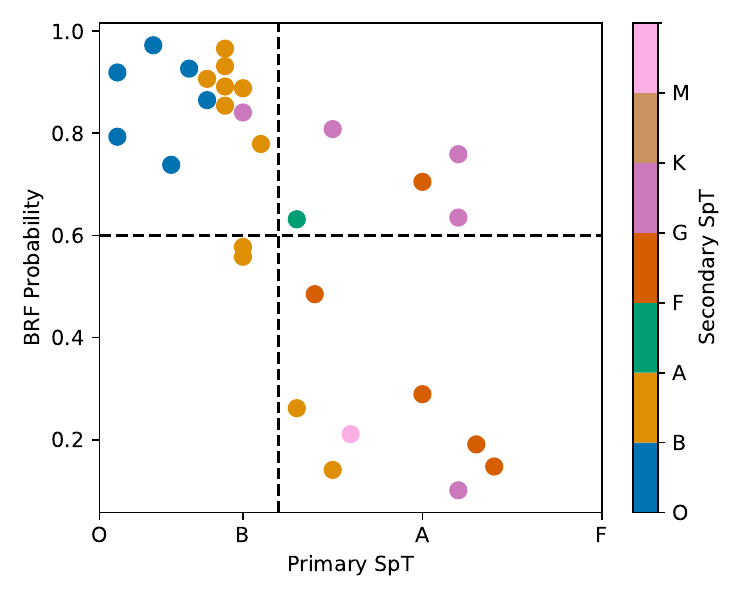}
\caption{BRF probability of binary systems as a function of the primary's spectral type. The spectral type of the secondary components is indicated by different colors.}
\label{fig:binaries}
\end{figure}

Figure \ref{fig:binaries} shows the spectral type of the primary stars against the BRF probability, with the spectral type of the secondary indicated by color. Systems with mass ratio ($q \sim 1$) have similar spectral types and are therefore expected to exhibit similar colors. We observe that most systems with a massive primary and q$\sim$ 1 are identified by our methodology as massive indeed. On the other hand, the recovery efficiency is not $100\%$ as two systems with a B primary and a close to 1 mass ratio lie just below threshold probability. With this small test we can assess that, although we introduced a bias removing multiples from the training set, this does not seem to have a very strong effect (baring the small numbers of the experiment) in correctly predicting multiples with a massive primary.

\subsection{Homogeneous validation dataset: LAMOST spectroscopy}\label{sec:LAMOST spectroscopy}

We searched for spectral information in the Large Sky Area Multi-Object Fiber Spectroscopic Telescope (LAMOST, \citealt{2012RAA....12..723Z}) to validate the classifier performance. This survey can obtain 4 000 spectra in one exposure for a limiting magnitude of 19 in r band and a spectral resolving power $R \sim 1800$ around the g band. It has a Field of View (FoV) larger than 20 square degrees and a fiber scale of 3.3\arcsec on the focal plane. The last public data release (DR10) \footnote{\href{http://www.lamost.org/dr10/}{http://www.lamost.org/dr10/}} contained low-resolution spectra for 11100139 stars, covering approximately $-10^\circ < \text{DEC} < 89^\circ$, making it a valuable database for large-scale spectroscopic studies.

\begin{figure}[htpb]
\centering
\includegraphics[scale=0.6]{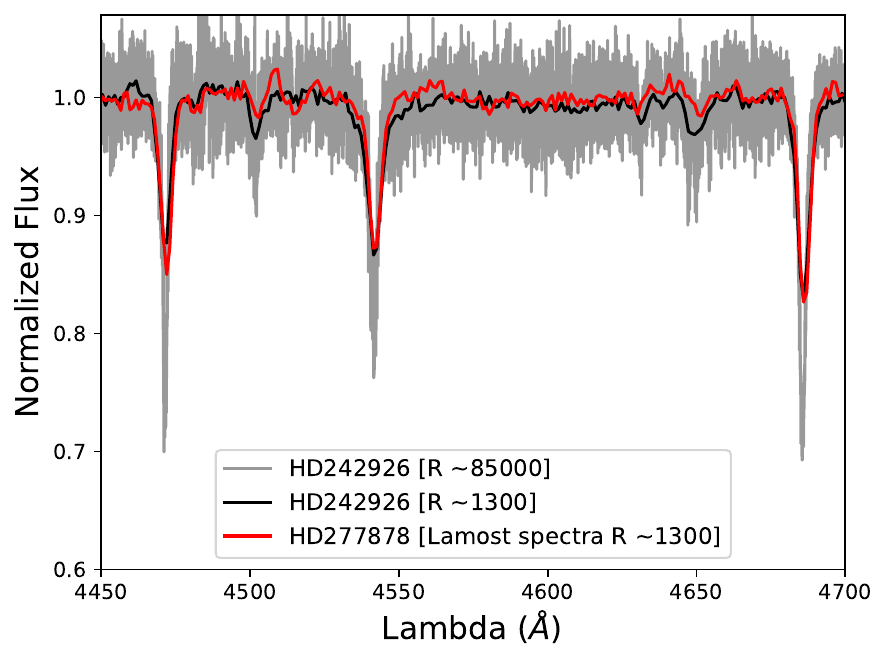}
\caption{The massive standard star HD 242926, classified as O7Vz, is shown at high resolution (grayline) and degraded to LAMOST resolution (black line), in comparison to the O7V((f))z star HD 277878, observed by LAMOST (red line).}
\label{fig:Standard_degraded}
\end{figure}

\subsubsection{Automatic spectral classification}

We found 1066 spectra in LAMOST within a 5\arcsec radius for the moderate to high probability massive star candidates, considering all matches. To enhance the precision of our selection, we only considered counterparts located at the intersection of the initial 5\arcsec radius and a 5\arcsec wide band along the star's proper motion direction. We discarded 106 spectra that showed anomalous offsets inconsistent with the expected proper motion direction. The final sample contained 880 spectra, from which we identified 585 unique sources. For each source with multiple observations, we retained the spectrum with the highest signal-to-noise ratio (SNR).

To derive spectral types for this sample, we developed an automatic methodology for classifying the low-resolution LAMOST spectra, based on a ground-truth sample of 50 massive standard stars of luminosity class V with spectral types O, B, and A collected from \cite{1982A&AS...50..199D, sota2011, 2016ApJS..224....4M,2018A&A...616A.135M}. The ground-truth spectra were obtained using the High-Efficiency and high-Resolution Mercator Echelle Spectrograph (HERMES; \citeauthor{2011A&A...526A..69R} 2011) and  the Fiber-fed Extended Range Optical Spectrograph (FEROS; \citeauthor{1999Msngr..95....8K} 1999), with resolutions of 85000 and 48000, respectively.

\begin{figure}[htpb]
\centering
\includegraphics[scale=0.8]{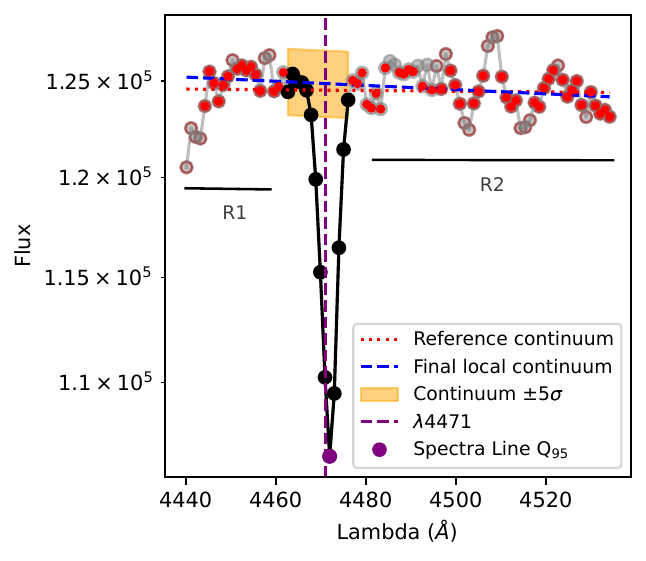}
\caption{Automatic method to detect the continuum of the \ion{He}{i} $\lambda$4471 spectral line. In gray, the points used to estimate the reference continuum. In red, full circles represent the points used to estimate the final local continuum. The $Q_{95}$ (in purple) is used to determine if the peak or absorption exceeds 5$\sigma$ above the continuum.}
\label{fig:DetectionLine}
\end{figure}
\begin{figure*}[htpb]
  \begin{center}
    \includegraphics[scale=0.65]{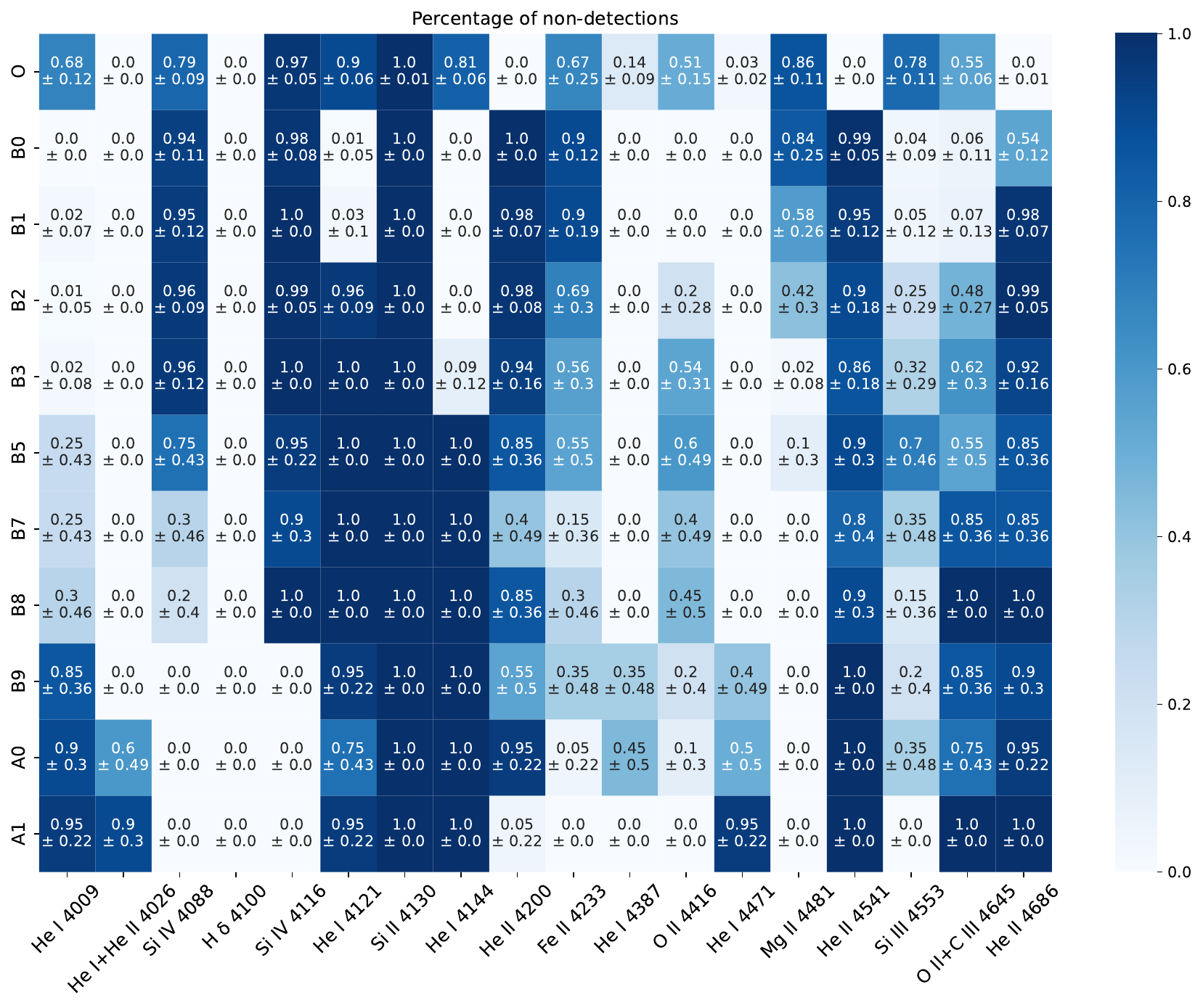}
  \end{center}
  \caption{Percentage of non-detected lines across different spectral types of massive standard stars. For each iteration, we used a random sample of SNR values from the distribution of SNRs in LAMOST candidates. We present the mean detection rate and the standard deviation for 20 iterations.}
  \label{fig:CM_lamost}
\end{figure*}

We degraded the high-resolution spectra of standard dwarfs to match LAMOST's observational capabilities using the \textit{specutils} package \citep{nicholas_earl_2024_14042033}\footnote{\url{https://specutils.readthedocs.io/en/stable}}. We applied convolution based smoothing to the spectrum using a Gaussian kernel, reducing the resolution to R$\sim$1300, which is the expected resolution for LAMOST at \ion{He}{i} \(\lambda 4471\) (see Figure 2 of \cite{2006SPIE.6269E..0MZ}), an important line for distinguishing O and B stars. We resampled the flux to the LAMOST dispersion grid while preserving the integrated flux and added normal noise to compensate for the increase in SNR caused by the resolution reduction. We used a Wasserstein distance to compare the SNR distributions before and after adding noise. The distance decreased from 0.008 to 0.001, making the SNR of both distributions more similar. In Fig. \ref{fig:Standard_degraded}, we show the spectrum of the massive standard star HD 242926, classified as O7Vz \citep{sota2011}, as observed with HERMES, along with its degraded version at LAMOST resolution. For comparison, we also include the LAMOST spectrum of the O7V((f))z star HD 277878 \citep{2019MNRAS.484.5578R}.

We selected 18 spectral lines used in \cite{2022A&A...657A..62K} and tested which lines were detected in the standard dwarfs with low resolution. For the detection of the lines, we combined the methodology presented in \cite{2011A&A...536A..63B} and \cite{2022A&A...657A..62K}. The first approach detected the continuum of the line in a completely automatic way, while the second one used static spectral windows, taking into account the surroundings of each line.

We present an example of the detection of the \ion{He}{i} \(\lambda 4471\) line in Fig. \ref{fig:DetectionLine}, the same procedure was applied to each line. We defined two fixed spectral regions (R1 and R2), in terms of central wavelength and width, to model the continuum, following \cite{2022A&A...657A..62K}, and a third one to trace the line (hereafter, the "line region"). An initial estimate of the continuum was derived from a linear fit to the R1 and R2 regions. This continuum was then refined by applying a one-sigma clipping. Once the final local continuum was identified, we assessed the significance of a detection by comparing the flux within the line region to the dispersion of the continuum. We subtracted the continuum and considered a detection whenever the peak of the line (defined as the 95 percentile of the flux distribution) was above 5 sigma.

We calculated the equivalent width (EW) of the line if it was detected. We considered that the line follows a Gaussian profile and calculated the associated $\sigma$. We performed three types of line fitting to characterize different scenarios: isolated line, blended line, and core emission line using a single Gaussian fit, a double Gaussian fit, and double inverted Gaussian fit, respectively. We selected the best fit using the Bayesian Information Criterion (BIC). This criterion is useful for model selection, giving priority to models with fewer parameters.

The line region was defined as $\pm4\sigma$ around the center of the fitting model. We found that this threshold provided the best separation between the \ion{He}{i} $\lambda$4471 and \ion{Mg}{ii} $\lambda$4481 lines in early-type stars when these lines are not blended. For consistency, the same criterion was applied to all other lines. Finally, we calculated the EW over the measured flux (not the fitted line), within the line region.

We applied the automatic methodology to characterize line detection in low-resolution spectra of dwarf massive standard stars. For a given spectral line, its detection can be influenced by the SNR of the spectra. To quantify the impact of SNR on line detection, we selected SNR values sampled from the distribution of moderate- to high-probability candidates and assessed the corresponding detections in the low-resolution spectra of massive dwarf standard stars.

We performed 20 iterations, each using a different SNR value. In each experiment, we calculated the detection rate for all lines across the sample. The resulting matrix, shown in Fig.~\ref{fig:CM_lamost}, reports the mean and standard deviation of the detection rates for each spectral type and line. This results are highly compatible with the results from \cite{2022A&A...657A..62K}. The \ion{He}{i} lines are clearly identified in B-type stars, while the \ion{He}{ii} lines are systematically observed in O-type stars. The Si lines remain challenging to detect in both works.

From this matrix, we selected the following lines for classification: \ion{He}{i} $\lambda$4471, \ion{Mg}{ii} $\lambda$4481, \ion{He}{i} $\lambda$4144, \ion{He}{ii} $\lambda$4541, \ion{He}{ii} $\lambda$4686, \ion{Fe}{ii} $\lambda$4233, \ion{He}{i} + \ion{He}{ii} $\lambda$4026. These lines are consistent with the expected spectral types, and based on them, we created an empirical decision path to classify the LAMOST spectra. It is important to note that we are only able to spectroscopically confirm stars of type B3 or earlier, due to the difficulty of detecting silicon lines at this resolution using an automatic approach, as illustrated in Fig.~\ref{fig:CM_lamost} and in agreement with \cite{2022A&A...657A..62K}.

Figure \ref{fig:decision_clasificacion} summarizes the empirical decision path using automated line detection to classify O and early B stars. The first selection focused on \ion{He}{ii} \(\lambda 4686\) and \ion{He}{ii} \(\lambda 4541\), as these lines were consistently present in O stars. To minimize false positives due to the SNR in B-type stars, we applied a strict detection criterion for both lines. We then used the detection and EW ratio of \ion{He}{i} \(\lambda 4471\) and \ion{Mg}{ii} \(\lambda 4481\), which is useful to classify helium-normal stars later than B3 \citep{2009ssc..book.....G}, with a ratio near 1 for B6-B7 stars \citep{2020ApJ...894....5R}. The \ion{Mg}{ii} \(\lambda 4481\) line was consistently detected from B3 to A1, while \ion{He}{i} \(\lambda 4471\) became undetectable from B9 onward and was almost always absent in A1 stars. We combined the detection of \ion{He}{i} + \ion{He}{ii} \(\lambda 4026\), which is always present in O and B stars, with the detection of \ion{He}{i} \(\lambda 4144\), primarily observed up to B3. For stars later than A1, we used the \ion{Fe}{ii} \(\lambda 4233\) line, characteristic of A1 stars, and kept the designation "later" since we did not include spectra of standard stars of later types.

\begin{figure}[htpb]
\centering
\includegraphics[scale=0.45]{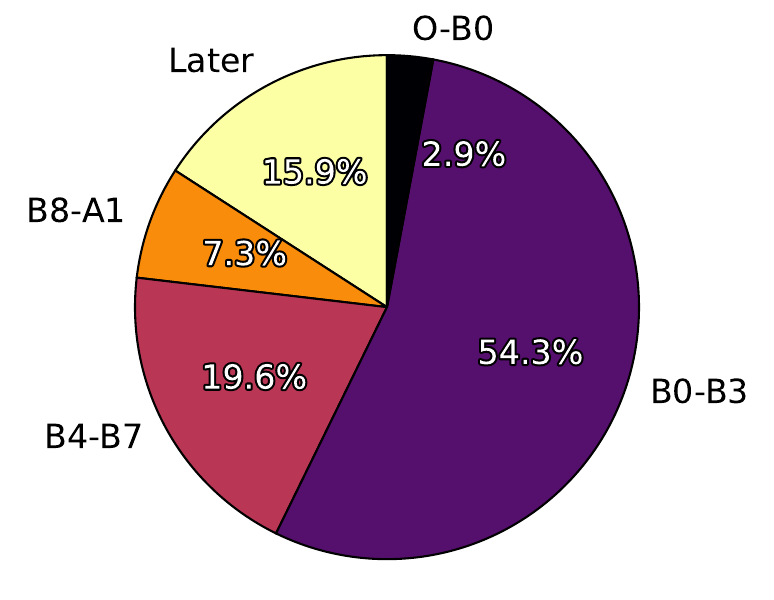}
\caption{Spectroscopic classification percentages of moderate to high probability massive star candidates in the LAMOST sample, determined using the automatic methodology and empirical decision path.}
\label{fig:pie_chart_lamost}
\end{figure}

\subsubsection{Estimated spectral types for the validation data-set}
\begin{figure}
\centering
\includegraphics[scale=0.5]{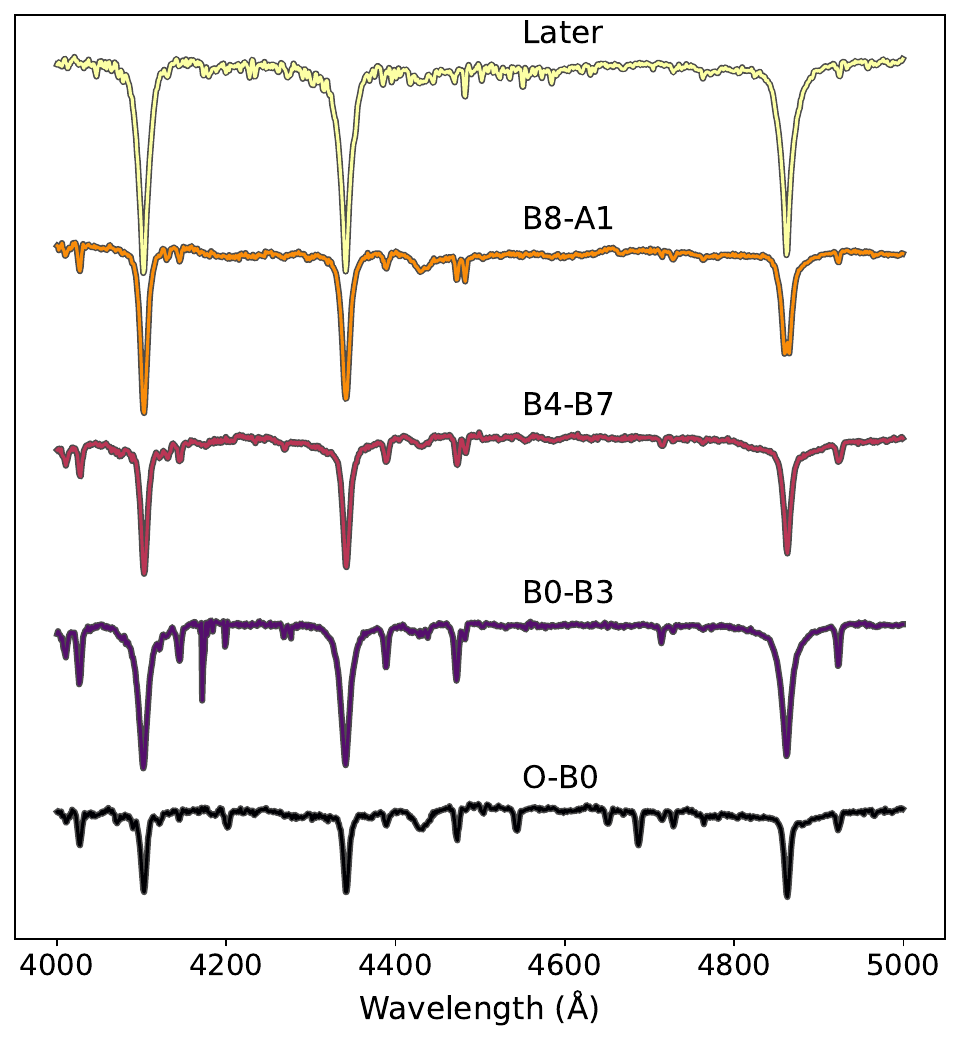}
\caption{Examples of spectra from the LAMOST sample, classified into spectral type groups, with an SNR in the g band detection in SDSS of approximately 250.}
\label{fig:example_classification_lamost}
\end{figure}

We applied the automatic methodology and empirical decision path to classify the LAMOST spectra of moderate to high probability massive star candidates. Fig. \ref{fig:pie_chart_lamost} presents the classification percentages, while Fig. \ref{fig:example_classification_lamost} illustrates examples of spectra classified with an SNR in the g band detection in SDSS of approximately 250.  We classified $57.2\%$ of the sample as massive stars. Approximately $19.6\%$ were misclassified as mid B-type stars (B4–B7). The results could improve by considering the luminosity class; however, developing an automated methodology for luminosity classification is beyond the scope of this work. We refer to \cite{2024A&A...690A.176N} for the standard classification of B stars. Additionally, $16\%$ of the sample corresponds to stars later than type A, representing the lower limit of the model's misclassification.  

\begin{figure}[htpb]
\centering
\includegraphics[scale=0.4]{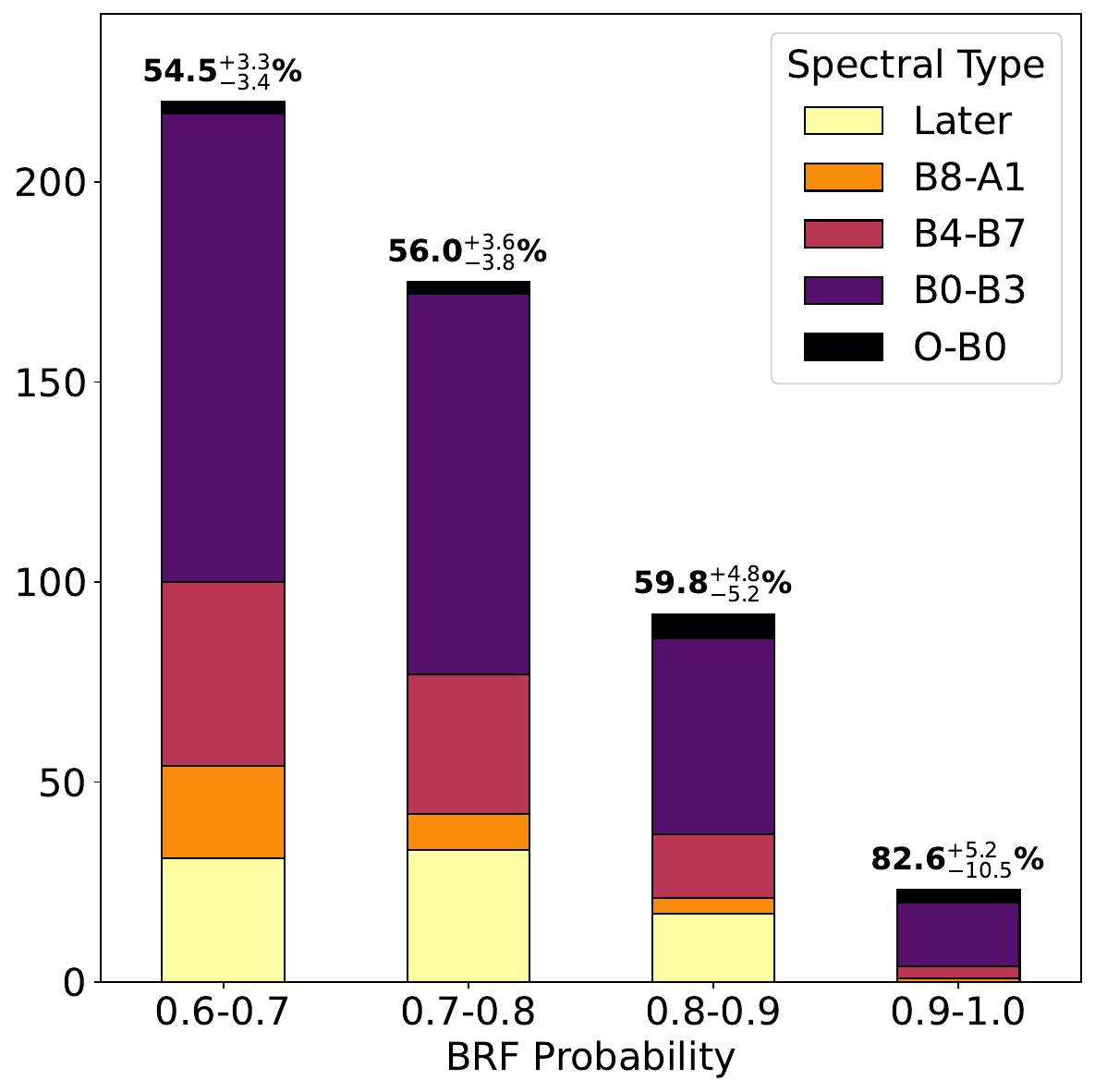}
\caption{Number of stars classified in different spectral type groups, divided by the probability given by the BRF. The top of each bar shows the detection rate of massive stars.}
\label{fig:detection_rate}
\end{figure}

Figure \ref{fig:detection_rate} illustrates the number of stars classified into different spectral type groups, separated by the probability assigned by the BRF. At the top of each bar, we present the detection rate of massive stars, calculated as the percentage of the O-B0 and B0-B3 groups relative to all other spectral types within each probability interval. The detection rates align with the expected precision of $0.564 \pm 0.009 $ for detecting B3 or earlier spectral types, as shown in Table \ref{table:metrics}. We observe that the detection rate increases with higher probabilities, while the total number of classified stars decreases. This reduction in moderate to high probability candidates is due to the selection function of the LAMOST survey, which does not cover the Galactic center, where a higher concentration of massive star candidates is expected.

\subsection{Comparison with state-of-the-art massive star data bases and catalogs} \label{Sec:Discussions}

We selected dedicated studies focused on the spectroscopic classification of massive stars to evaluate the performance of our BRF algorithm. Specifically, we examined the IACOB spectroscopic database \citep{IACOB2011}, GOSSS \citep{Goss2013}, \cite{chini2012}, \cite{Li_2021}, and the ALS compilation \citep{2021MNRAS.504.2968P}. Within a 2\arcsec search radius, we identified 124, 36, 17, 29, and 831 stars, respectively, that were not included in our training or test sets.

\begin{figure}[htpb]
\centering
\includegraphics[scale=0.5]{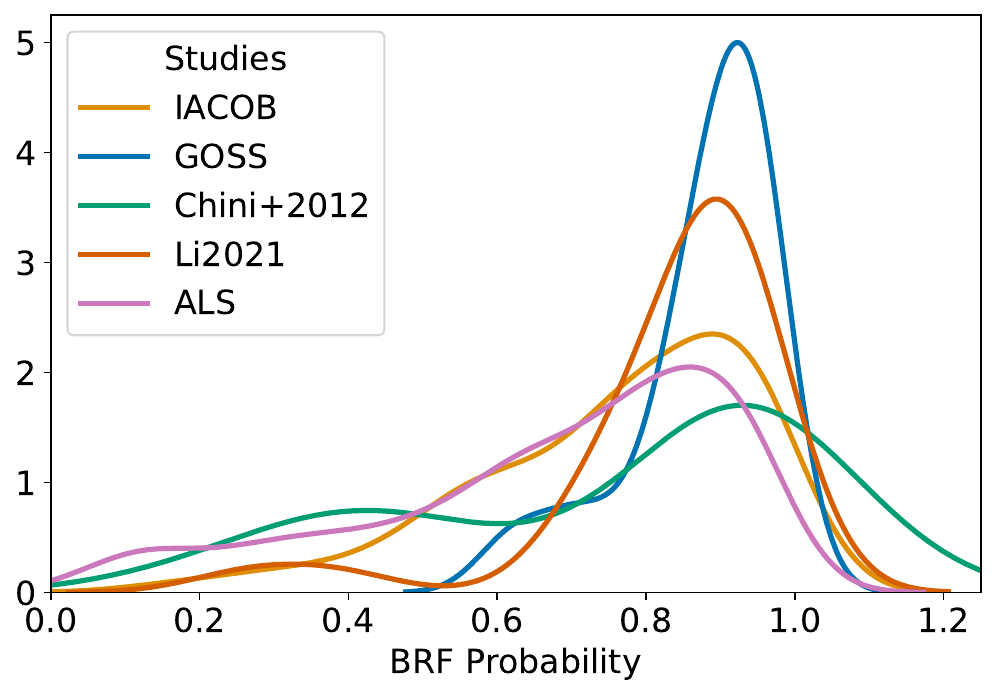}
\caption{Probability distribution of the BRF for dedicated studies of massive stars.}
\label{fig:massive_studies}
\end{figure}

Figure \ref{fig:massive_studies} presents the probability distribution for different studies. The GOSSS and \cite{Li_2021} samples are more concentrated at high probabilities, as they consist exclusively of O-type stars. In the GOSSS sample, we classified all stars with a probability greater than 0.5, whereas in the \cite{Li_2021} sample, we misclassified two O9 stars. The comparison with the \cite{chini2012}, showed the poorest classifier performance, with most misclassifications resulting from intermediate B-type stars present in this sample. Among the five misclassified stars, three were B4 V stars, one was a B2 V star, and one was a B1 III star. In the IACOB and ALS samples, we classified approximately 10\% and 20\% of the stars, respectively, with probabilities below 0.5. However, not all stars in these studies had luminosity classifications, and some belonged to later B spectral types. In general, the probabilities were high, with a missed detection rate consistent with the algorithm's expected metrics.

\section{Summary and Conclusions} \label{sec:summary}
In this study, we present a moderate to high-probability catalog of massive star candidates with estimated spectral types B2 or earlier in the Milky Way. We defined our parental catalog using stars with spectral type O to A with astrometry and photometry from Gaia DR3, combined with photometry from 2MASS. We defined a subsample spectroscopically confirmed from the literature to serve as a source of labels. Since the labeled sample is biased toward brighter stars, we restricted our sample to $G < 12$~mag. In addition, we discarded possible outliers by discarding objects with any feature outside of the 1-99 percentile range.

A BRF classifier was trained on the encoded labeled data, simplifying the taxonomy to ``B2 or earlier" or ``later than B2" spectral types. The trained model was used to predict spectral types in this binary fashion on the unlabeled sample. The performance of our classifier shows high completeness of $0.8$ and a purity ranging from $0.51 \pm 0.02$ for probabilities between 0.6 and 0.7, up to $0.85 \pm 0.05$ for the 0.9–1.0 range.

We mined spectroscopic archives for data on our moderate to high-probability candidates (estimated ``B2 or earlier spectral type" with $P > 0.6$). The most homogeneous subsample obtained consisted of 534 objects with available spectra from the LAMOST survey. We developed an automatic methodology to detect the presence and characteristics of spectral lines on a sample of dwarf massive standard stars. Based on the extracted line(s) information, we designed a decision path capable of differentiating stars with spectral types O--B0, B0--B3, B4--B7, B8--A1, and later than A1. Applying this scheme to the 534 validation set, we obtained a confirmation rate of $\sim$54\%. We further explored the dependence of this confirmation rate on the probability cut form the model, reaching 83\% for $P>0.9$. From the validation set (534 source), to the best of our knowledge 107 are completely new discoveries.

As a result of our work, we obtained a catalog of $\sim$9000 moderate to high probability $(P>0.6)$ ``B2 or earlier" bright ($G<12$ mag) stars. From those, 5158 are not present in known massive stars databases (or general databases as Simbad). The remaining, 3931 were not present in our homogeneous set of encoded labels but were reported in other works in the literature. The number of new candidates presented in this work (5158) is therefore comparable to the currently known number of stars spectroscopically confirmed with spectral type B2 or earlier, 7500 and 5793 for G < 12 mag respectively \citep{IACOB2011,chini2012,Goss2013,2014yCat....1.2023S,2021MNRAS.504.2968P,Li_2021}. Assuming our metrics hold for the full sample, we expect to increase the known population of massive stars by approximately 50\%.

\begin{acknowledgements}
We thank the anonymous referee for the helpful comments that improved the quality of this work. We acknowledge the use of TOPCAT \citep{2005ASPC..347...29T} and the following Python libraries used in this research: pandas \citep{reback2020pandas}, matplotlib \citep{Hunter:2007}, numpy \citep{harris2020array}, ray \citep{2017arXiv171205889M}, scikit-learn \citep{scikit-learn}, imblearn \citep{JMLR:v18:16-365}, astropy \citep{astropy:2013,astropy:2018,astropy:2022}, scipy \citep{2020SciPy-NMeth}, and specutils \citep{nicholas_earl_2024_14042033}. We acknowledge support from the National Agency for Research and Development (ANID) through the Scholarship Program DOCTORADO BECAS CHILE/2021 - 21211323. A.B acknowledges support from the Deutsche Forschungsgemeinschaft (DFG, German Research Foundation) under Germany's Excellence Strategy – EXC 2094 – 390783311. Based on observations made with the Mercator Telescope, operated on the island of La Palma by the Flemish Community, at the Spanish Observatorio del Roque de los Muchachos of the Instituto de Astrofísica de Canarias. Based on observations obtained with the HERMES spectrograph, which is supported by the Research Foundation - Flanders (FWO), Belgium, the Research Council of KU Leuven, Belgium, the Fonds National de la Recherche Scientifique (F.R.S.-FNRS), Belgium, the Royal Observatory of Belgium, the Observatoire de Genève, Switzerland and the Thüringer Landessternwarte Tautenburg, Germany. Guoshoujing Telescope (the Large Sky Area Multi-Object Fiber Spectroscopic Telescope LAMOST) is a National Major Scientific Project built by the Chinese Academy of Sciences. Funding for the project has been provided by the National Development and Reform Commission. LAMOST is operated and managed by the National Astronomical Observatories, Chinese Academy of Sciences. This work has made use of data from the European Space Agency (ESA) mission
{\it Gaia} (\url{https://www.cosmos.esa.int/gaia}), processed by the {\it Gaia}
Data Processing and Analysis Consortium (DPAC,
\url{https://www.cosmos.esa.int/web/gaia/dpac/consortium}). Funding for the DPAC
has been provided by national institutions, in particular the institutions
participating in the {\it Gaia} Multilateral Agreement.
\end{acknowledgements}

%
%
\bibliographystyle{aa}
\bibliography{bibliography.bib}

\begin{thebibliography}{81}
\expandafter\ifx\csname natexlab\endcsname\relax\def\natexlab#1{#1}\fi

\bibitem[{{Astropy Collaboration} {et~al.}(2022){Astropy Collaboration},
  {Price-Whelan}, {Lim}, {Earl}, {Starkman}, {Bradley}, {Shupe}, {Patil},
  {Corrales}, {Brasseur}, {N{"o}the}, {Donath}, {Tollerud}, {Morris},
  {Ginsburg}, {Vaher}, {Weaver}, {Tocknell}, {Jamieson}, {van Kerkwijk},
  {Robitaille}, {Merry}, {Bachetti}, {G{"u}nther}, {Aldcroft},
  {Alvarado-Montes}, {Archibald}, {B{'o}di}, {Bapat}, {Barentsen}, {Baz{'a}n},
  {Biswas}, {Boquien}, {Burke}, {Cara}, {Cara}, {Conroy}, {Conseil}, {Craig},
  {Cross}, {Cruz}, {D'Eugenio}, {Dencheva}, {Devillepoix}, {Dietrich},
  {Eigenbrot}, {Erben}, {Ferreira}, {Foreman-Mackey}, {Fox}, {Freij}, {Garg},
  {Geda}, {Glattly}, {Gondhalekar}, {Gordon}, {Grant}, {Greenfield}, {Groener},
  {Guest}, {Gurovich}, {Handberg}, {Hart}, {Hatfield-Dodds}, {Homeier},
  {Hosseinzadeh}, {Jenness}, {Jones}, {Joseph}, {Kalmbach}, {Karamehmetoglu},
  {Ka{l}uszy{'n}ski}, {Kelley}, {Kern}, {Kerzendorf}, {Koch}, {Kulumani},
  {Lee}, {Ly}, {Ma}, {MacBride}, {Maljaars}, {Muna}, {Murphy}, {Norman},
  {O'Steen}, {Oman}, {Pacifici}, {Pascual}, {Pascual-Granado}, {Patil},
  {Perren}, {Pickering}, {Rastogi}, {Roulston}, {Ryan}, {Rykoff}, {Sabater},
  {Sakurikar}, {Salgado}, {Sanghi}, {Saunders}, {Savchenko}, {Schwardt},
  {Seifert-Eckert}, {Shih}, {Jain}, {Shukla}, {Sick}, {Simpson},
  {Singanamalla}, {Singer}, {Singhal}, {Sinha}, {Sip{H{o}}cz}, {Spitler},
  {Stansby}, {Streicher}, {{{S}}umak}, {Swinbank}, {Taranu}, {Tewary},
  {Tremblay}, {Val-Borro}, {Van Kooten}, {Vasovi{'c}}, {Verma}, {de Miranda
  Cardoso}, {Williams}, {Wilson}, {Winkel}, {Wood-Vasey}, {Xue}, {Yoachim},
  {Zhang}, {Zonca}, \& {Astropy Project Contributors}}]{astropy:2022}
{Astropy Collaboration}, {Price-Whelan}, A.~M., {Lim}, P.~L., {et~al.} 2022,
  \apj, 935, 167

\bibitem[{{Astropy Collaboration} {et~al.}(2018){Astropy Collaboration},
  {Price-Whelan}, {Sip{\H{o}}cz}, {G{\"u}nther}, {Lim}, {Crawford}, {Conseil},
  {Shupe}, {Craig}, {Dencheva}, {Ginsburg}, {Vand erPlas}, {Bradley},
  {P{\'e}rez-Su{\'a}rez}, {de Val-Borro}, {Aldcroft}, {Cruz}, {Robitaille},
  {Tollerud}, {Ardelean}, {Babej}, {Bach}, {Bachetti}, {Bakanov}, {Bamford},
  {Barentsen}, {Barmby}, {Baumbach}, {Berry}, {Biscani}, {Boquien}, {Bostroem},
  {Bouma}, {Brammer}, {Bray}, {Breytenbach}, {Buddelmeijer}, {Burke},
  {Calderone}, {Cano Rodr{\'\i}guez}, {Cara}, {Cardoso}, {Cheedella}, {Copin},
  {Corrales}, {Crichton}, {D'Avella}, {Deil}, {Depagne}, {Dietrich}, {Donath},
  {Droettboom}, {Earl}, {Erben}, {Fabbro}, {Ferreira}, {Finethy}, {Fox},
  {Garrison}, {Gibbons}, {Goldstein}, {Gommers}, {Greco}, {Greenfield},
  {Groener}, {Grollier}, {Hagen}, {Hirst}, {Homeier}, {Horton}, {Hosseinzadeh},
  {Hu}, {Hunkeler}, {Ivezi{\'c}}, {Jain}, {Jenness}, {Kanarek}, {Kendrew},
  {Kern}, {Kerzendorf}, {Khvalko}, {King}, {Kirkby}, {Kulkarni}, {Kumar},
  {Lee}, {Lenz}, {Littlefair}, {Ma}, {Macleod}, {Mastropietro}, {McCully},
  {Montagnac}, {Morris}, {Mueller}, {Mumford}, {Muna}, {Murphy}, {Nelson},
  {Nguyen}, {Ninan}, {N{\"o}the}, {Ogaz}, {Oh}, {Parejko}, {Parley}, {Pascual},
  {Patil}, {Patil}, {Plunkett}, {Prochaska}, {Rastogi}, {Reddy Janga},
  {Sabater}, {Sakurikar}, {Seifert}, {Sherbert}, {Sherwood-Taylor}, {Shih},
  {Sick}, {Silbiger}, {Singanamalla}, {Singer}, {Sladen}, {Sooley},
  {Sornarajah}, {Streicher}, {Teuben}, {Thomas}, {Tremblay}, {Turner},
  {Terr{\'o}n}, {van Kerkwijk}, {de la Vega}, {Watkins}, {Weaver}, {Whitmore},
  {Woillez}, {Zabalza}, \& {Astropy Contributors}}]{astropy:2018}
{Astropy Collaboration}, {Price-Whelan}, A.~M., {Sip{\H{o}}cz}, B.~M., {et~al.}
  2018, \aj, 156, 123

\bibitem[{{Astropy Collaboration} {et~al.}(2013){Astropy Collaboration},
  {Robitaille}, {Tollerud}, {Greenfield}, {Droettboom}, {Bray}, {Aldcroft},
  {Davis}, {Ginsburg}, {Price-Whelan}, {Kerzendorf}, {Conley}, {Crighton},
  {Barbary}, {Muna}, {Ferguson}, {Grollier}, {Parikh}, {Nair}, {Unther},
  {Deil}, {Woillez}, {Conseil}, {Kramer}, {Turner}, {Singer}, {Fox}, {Weaver},
  {Zabalza}, {Edwards}, {Azalee Bostroem}, {Burke}, {Casey}, {Crawford},
  {Dencheva}, {Ely}, {Jenness}, {Labrie}, {Lim}, {Pierfederici}, {Pontzen},
  {Ptak}, {Refsdal}, {Servillat}, \& {Streicher}}]{astropy:2013}
{Astropy Collaboration}, {Robitaille}, T.~P., {Tollerud}, E.~J., {et~al.} 2013,
  \aap, 558, A33

\bibitem[{{Bally} \& {Zinnecker}(2005)}]{Bally2005}
{Bally}, J. \& {Zinnecker}, H. 2005, \aj, 129, 2281

\bibitem[{{Barbá} {et~al.}(2017){Barbá}, {Gamen}, {Arias}, \&
  {Morrell}}]{Barba2017}
{Barbá}, R.~H., {Gamen}, R., {Arias}, J.~I., \& {Morrell}, N.~I. 2017, in {IAU
  Symposium}, Vol. 329, {The Lives and Death-Throes of Massive Stars}, ed.
  J.~J. {Eldridge}, J.~C. {Bray}, L.~A.~S. {McClelland}, \& L.~{Xiao}, 89–96

\bibitem[{{Bastian} {et~al.}(2010){Bastian}, {Covey}, \& {Meyer}}]{Bastian2010}
{Bastian}, N., {Covey}, K.~R., \& {Meyer}, M.~R. 2010, \araa, 48, 339

\bibitem[{{Bayo} {et~al.}(2011){Bayo}, {Barrado}, {Stauffer},
  {Morales-Calder{\'o}n}, {Melo}, {Hu{\'e}lamo}, {Bouy}, {Stelzer}, {Tamura},
  \& {Jayawardhana}}]{2011A&A...536A..63B}
{Bayo}, A., {Barrado}, D., {Stauffer}, J., {et~al.} 2011, \aap, 536, A63

\bibitem[{{Bayo} {et~al.}(2008){Bayo}, {Rodrigo}, {Barrado Y Navascu{\'e}s},
  {Solano}, {Guti{\'e}rrez}, {Morales-Calder{\'o}n}, \& {Allard}}]{Bayo2008}
{Bayo}, A., {Rodrigo}, C., {Barrado Y Navascu{\'e}s}, D., {et~al.} 2008, \aap,
  492, 277

\bibitem[{{Belokurov} {et~al.}(2020){Belokurov}, {Penoyre}, {Oh}, {Iorio},
  {Hodgkin}, {Evans}, {Everall}, {Koposov}, {Tout}, {Izzard}, {Clarke}, \&
  {Brown}}]{2020MNRAS.496.1922B}
{Belokurov}, V., {Penoyre}, Z., {Oh}, S., {et~al.} 2020, \mnras, 496, 1922

\bibitem[{{Breiman}(2001)}]{2001MachL..45....5B}
{Breiman}, L. 2001, Machine Learning, 45, 5

\bibitem[{{Cervi{\~n}o} {et~al.}(2013{\natexlab{a}}){Cervi{\~n}o},
  {Rom{\'a}n-Z{\'u}{\~n}iga}, {Bayo}, {Luridiana}, {S{\'a}nchez}, \&
  {P{\'e}rez}}]{cervino2013b}
{Cervi{\~n}o}, M., {Rom{\'a}n-Z{\'u}{\~n}iga}, C., {Bayo}, A., {et~al.}
  2013{\natexlab{a}}, \aap, 553, A32

\bibitem[{{Cervi{\~n}o} {et~al.}(2013{\natexlab{b}}){Cervi{\~n}o},
  {Rom{\'a}n-Z{\'u}{\~n}iga}, {Luridiana}, {Bayo}, {S{\'a}nchez}, \&
  {P{\'e}rez}}]{cervino2013a}
{Cervi{\~n}o}, M., {Rom{\'a}n-Z{\'u}{\~n}iga}, C., {Luridiana}, V., {et~al.}
  2013{\natexlab{b}}, \aap, 553, A31

\bibitem[{Chawla {et~al.}(2002)Chawla, Bowyer, Hall, \&
  Kegelmeyer}]{chawla2002smote}
Chawla, N.~V., Bowyer, K.~W., Hall, L.~O., \& Kegelmeyer, W.~P. 2002, Journal
  of artificial intelligence research, 16, 321

\bibitem[{Chen {et~al.}(2004)Chen, Liaw, Breiman, {et~al.}}]{chen2004using}
Chen, C., Liaw, A., Breiman, L., {et~al.} 2004, University of California,
  Berkeley, 110, 24

\bibitem[{{Chini} {et~al.}(2012){Chini}, {Hoffmeister}, {Nasseri}, {Stahl}, \&
  {Zinnecker}}]{chini2012}
{Chini}, R., {Hoffmeister}, V.~H., {Nasseri}, A., {Stahl}, O., \& {Zinnecker},
  H. 2012, \mnras, 424, 1925

\bibitem[{{de Jong} {et~al.}(2019){de Jong}, {Agertz}, {Berbel}, {Aird},
  {Alexander}, {Amarsi}, {Anders}, {Andrae}, {Ansarinejad}, {Ansorge},
  {Antilogus}, {Anwand-Heerwart}, {Arentsen}, {Arnadottir}, {Asplund}, {Auger},
  {Azais}, {Baade}, {Baker}, {Baker}, {Balbinot}, {Baldry}, {Banerji},
  {Barden}, {Barklem}, {Barth{\'e}l{\'e}my-Mazot}, {Battistini}, {Bauer},
  {Bell}, {Bellido-Tirado}, {Bellstedt}, {Belokurov}, {Bensby}, {Bergemann},
  {Bestenlehner}, {Bielby}, {Bilicki}, {Blake}, {Bland-Hawthorn}, {Boeche},
  {Boland}, {Boller}, {Bongard}, {Bongiorno}, {Bonifacio}, {Boudon}, {Brooks},
  {Brown}, {Brown}, {Br{\"u}ggen}, {Brynnel}, {Brzeski}, {Buchert},
  {Buschkamp}, {Caffau}, {Caillier}, {Carrick}, {Casagrande}, {Case}, {Casey},
  {Cesarini}, {Cescutti}, {Chapuis}, {Chiappini}, {Childress}, {Christlieb},
  {Church}, {Cioni}, {Cluver}, {Colless}, {Collett}, {Comparat}, {Cooper},
  {Couch}, {Courbin}, {Croom}, {Croton}, {Daguis{\'e}}, {Dalton}, {Davies},
  {Davis}, {de Laverny}, {Deason}, {Dionies}, {Disseau}, {Doel}, {D{\"o}scher},
  {Driver}, {Dwelly}, {Eckert}, {Edge}, {Edvardsson}, {Youssoufi}, {Elhaddad},
  {Enke}, {Erfanianfar}, {Farrell}, {Fechner}, {Feiz}, {Feltzing}, {Ferreras},
  {Feuerstein}, {Feuillet}, {Finoguenov}, {Ford}, {Fotopoulou}, {Fouesneau},
  {Frenk}, {Frey}, {Gaessler}, {Geier}, {Gentile Fusillo}, {Gerhard},
  {Giannantonio}, {Giannone}, {Gibson}, {Gillingham},
  {Gonz{\'a}lez-Fern{\'a}ndez}, {Gonzalez-Solares}, {Gottloeber}, {Gould},
  {Grebel}, {Gueguen}, {Guiglion}, {Haehnelt}, {Hahn}, {Hansen}, {Hartman},
  {Hauptner}, {Hawkins}, {Haynes}, {Haynes}, {Heiter}, {Helmi}, {Aguayo},
  {Hewett}, {Hinton}, {Hobbs}, {Hoenig}, {Hofman}, {Hook}, {Hopgood},
  {Hopkins}, {Hourihane}, {Howes}, {Howlett}, {Huet}, {Irwin}, {Iwert},
  {Jablonka}, {Jahn}, {Jahnke}, {Jarno}, {Jin}, {Jofre}, {Johl}, {Jones},
  {J{\"o}nsson}, {Jordan}, {Karovicova}, {Khalatyan}, {Kelz}, {Kennicutt},
  {King}, {Kitaura}, {Klar}, {Klauser}, {Kneib}, {Koch}, {Koposov},
  {Kordopatis}, {Korn}, {Kosmalski}, {Kotak}, {Kovalev}, {Kreckel}, {Kripak},
  {Krumpe}, {Kuijken}, {Kunder}, {Kushniruk}, {Lam}, {Lamer}, {Laurent},
  {Lawrence}, {Lehmitz}, {Lemasle}, {Lewis}, {Li}, {Lidman}, {Lind}, {Liske},
  {Lizon}, {Loveday}, {Ludwig}, {McDermid}, {Maguire}, {Mainieri}, {Mali}, \&
  {Mandel}}]{jong2019}
{de Jong}, R.~S., {Agertz}, O., {Berbel}, A.~A., {et~al.} 2019, The Messenger,
  175, 3

\bibitem[{de~Jong {et~al.}(2016)de~Jong, Barden, Bellido-Tirado, Brynnel, Frey,
  Giannone, Haynes, Johl, Phillips, Schnurr, Walcher, Winkler, Ansorge,
  Feltzing, McMahon, Baker, Caillier, Dwelly, Gaessler, Iwert, Mandel,
  Piskunov, Pragt, Walton, Bensby, Bergemann, Chiappini, Christlieb, Cioni,
  Driver, Finoguenov, Helmi, Irwin, Kitaura, Kneib, Liske, Merloni, Minchev,
  Richard, \& Starkenburg}]{Jong2016}
de~Jong, R.~S., Barden, S.~C., Bellido-Tirado, O., {et~al.} 2016, in
  Ground-based and Airborne Instrumentation for Astronomy VI, ed. C.~J. Evans,
  L.~Simard, \& H.~Takami, Vol. 9908, International Society for Optics and
  Photonics (SPIE), 99081O

\bibitem[{{Didelon}(1982)}]{1982A&AS...50..199D}
{Didelon}, P. 1982, \aaps, 50, 199

\bibitem[{{Dodd} {et~al.}(2024){Dodd}, {Oudmaijer}, {Radley}, {Vioque}, \&
  {Frost}}]{2024MNRAS.527.3076D}
{Dodd}, J.~M., {Oudmaijer}, R.~D., {Radley}, I.~C., {Vioque}, M., \& {Frost},
  A.~J. 2024, \mnras, 527, 3076

\bibitem[{Earl {et~al.}(2024)Earl, Tollerud, O'Steen, brechmos, Kerzendorf,
  Busko, shaileshahuja, D'Avella, Lim, Robitaille, Ginsburg, Homeier, Sipőcz,
  Averbukh, Tocknell, Cherinka, Ogaz, Geda, Conroy, Davies, Günther, Barbary,
  Foster, Droettboom, Nguyen, Bray, Casey, Cruz, Ferguson, \&
  Crawford}]{nicholas_earl_2024_14042033}
Earl, N., Tollerud, E., O'Steen, R., {et~al.} 2024, astropy/specutils: v1.19.0

\bibitem[{Eldridge \& Stanway(2022)}]{Eldridge2022}
Eldridge, J.~J. \& Stanway, E.~R. 2022, \araa, 60, 455–494

\bibitem[{{Gaia Collaboration} {et~al.}(2018){Gaia Collaboration}, {Brown},
  {Vallenari}, {Prusti}, {de Bruijne}, {Babusiaux}, {Bailer-Jones}, {Biermann},
  {Evans}, {Eyer}, {Jansen}, {Jordi}, {Klioner}, {Lammers}, {Lindegren},
  {Luri}, {Mignard}, {Panem}, {Pourbaix}, {Randich}, {Sartoretti}, {Siddiqui},
  {Soubiran}, {van Leeuwen}, {Walton}, {Arenou}, {Bastian}, {Cropper},
  {Drimmel}, {Katz}, {Lattanzi}, {Bakker}, {Cacciari}, {Casta{\~n}eda},
  {Chaoul}, {Cheek}, {De Angeli}, {Fabricius}, {Guerra}, {Holl}, {Masana},
  {Messineo}, {Mowlavi}, {Nienartowicz}, {Panuzzo}, {Portell}, {Riello},
  {Seabroke}, {Tanga}, {Th{\'e}venin}, {Gracia-Abril}, {Comoretto},
  {Garcia-Reinaldos}, {Teyssier}, {Altmann}, {Andrae}, {Audard},
  {Bellas-Velidis}, {Benson}, {Berthier}, {Blomme}, {Burgess}, {Busso},
  {Carry}, {Cellino}, {Clementini}, {Clotet}, {Creevey}, {Davidson}, {De
  Ridder}, {Delchambre}, {Dell'Oro}, {Ducourant},
  {Fern{\'a}ndez-Hern{\'a}ndez}, {Fouesneau}, {Fr{\'e}mat}, {Galluccio},
  {Garc{\'\i}a-Torres}, {Gonz{\'a}lez-N{\'u}{\~n}ez}, {Gonz{\'a}lez-Vidal},
  {Gosset}, {Guy}, {Halbwachs}, {Hambly}, {Harrison}, {Hern{\'a}ndez},
  {Hestroffer}, {Hodgkin}, {Hutton}, {Jasniewicz}, {Jean-Antoine-Piccolo},
  {Jordan}, {Korn}, {Krone-Martins}, {Lanzafame}, {Lebzelter}, {L{\"o}ffler},
  {Manteiga}, {Marrese}, {Mart{\'\i}n-Fleitas}, {Moitinho}, {Mora}, {Muinonen},
  {Osinde}, {Pancino}, {Pauwels}, {Petit}, {Recio-Blanco}, {Richards},
  {Rimoldini}, {Robin}, {Sarro}, {Siopis}, {Smith}, {Sozzetti}, {S{\"u}veges},
  {Torra}, {van Reeven}, {Abbas}, {Abreu Aramburu}, {Accart}, {Aerts},
  {Altavilla}, {{\'A}lvarez}, {Alvarez}, {Alves}, {Anderson}, {Andrei},
  {Anglada Varela}, {Antiche}, {Antoja}, {Arcay}, {Astraatmadja}, {Bach},
  {Baker}, {Balaguer-N{\'u}{\~n}ez}, {Balm}, {Barache}, {Barata}, {Barbato},
  {Barblan}, {Barklem}, {Barrado}, {Barros}, {Barstow}, {Bartholom{\'e}
  Mu{\~n}oz}, {Bassilana}, {Becciani}, {Bellazzini}, {Berihuete}, {Bertone},
  {Bianchi}, {Bienaym{\'e}}, {Blanco-Cuaresma}, {Boch}, {Boeche}, {Bombrun},
  {Borrachero}, {Bossini}, {Bouquillon}, {Bourda}, {Bragaglia}, {Bramante},
  {Breddels}, {Bressan}, {Brouillet}, {Br{\"u}semeister}, {Brugaletta},
  {Bucciarelli}, {Burlacu}, {Busonero}, {Butkevich}, {Buzzi}, {Caffau},
  {Cancelliere}, {Cannizzaro}, {Cantat-Gaudin}, {Carballo}, {Carlucci},
  {Carrasco}, {Casamiquela}, {Castellani}, {Castro-Ginard}, {Charlot},
  {Chemin}, {Chiavassa}, {Cocozza}, {Costigan}, {Cowell}, {Crifo}, {Crosta},
  {Crowley}, {Cuypers}, {Dafonte}, {Damerdji}, {Dapergolas}, {David}, {David},
  {de Laverny}, {De Luise}, {De March}, {de Martino}, {de Souza}, {de Torres},
  {Debosscher}, {del Pozo}, {Delbo}, {Delgado}, {Delgado}, {Di Matteo},
  {Diakite}, {Diener}, {Distefano}, {Dolding}, {Drazinos}, {Dur{\'a}n},
  {Edvardsson}, {Enke}, {Eriksson}, {Esquej}, {Eynard Bontemps}, {Fabre},
  {Fabrizio}, {Faigler}, {Falc{\~a}o}, {Farr{\`a}s Casas}, {Federici},
  {Fedorets}, {Fernique}, {Figueras}, {Filippi}, {Findeisen}, {Fonti},
  {Fraile}, {Fraser}, {Fr{\'e}zouls}, {Gai}, {Galleti}, {Garabato},
  {Garc{\'\i}a-Sedano}, {Garofalo}, {Garralda}, {Gavel}, {Gavras}, {Gerssen},
  {Geyer}, {Giacobbe}, {Gilmore}, {Girona}, {Giuffrida}, {Glass}, {Gomes},
  {Granvik}, {Gueguen}, {Guerrier}, {Guiraud}, {Guti{\'e}rrez-S{\'a}nchez},
  {Haigron}, {Hatzidimitriou}, {Hauser}, {Haywood}, {Heiter}, {Helmi}, {Heu},
  {Hilger}, {Hobbs}, {Hofmann}, {Holland}, {Huckle}, {Hypki}, {Icardi},
  {Jan{\ss}en}, {Jevardat de Fombelle}, {Jonker}, {Juh{\'a}sz}, {Julbe},
  {Karampelas}, {Kewley}, {Klar}, {Kochoska}, {Kohley}, {Kolenberg},
  {Kontizas}, {Kontizas}, {Koposov}, {Kordopatis}, {Kostrzewa-Rutkowska},
  {Koubsky}, {Lambert}, {Lanza}, {Lasne}, {Lavigne}, {Le Fustec}, {Le
  Poncin-Lafitte}, {Lebreton}, {Leccia}, {Leclerc}, {Lecoeur-Taibi},
  {Lenhardt}, {Leroux}, {Liao}, {Licata}, {Lindstr{\o}m}, {Lister}, {Livanou},
  {Lobel}, {L{\'o}pez}, {Managau}, {Mann}, {Mantelet}, {Marchal}, {Marchant},
  {Marconi}, {Marinoni}, {Marschalk{\'o}}, {Marshall}, {Martino}, {Marton},
  {Mary}, {Massari}, {Matijevi{\v{c}}}, {Mazeh}, {McMillan}, {Messina},
  {Michalik}, {Millar}, {Molina}, {Molinaro}, {Moln{\'a}r}, {Montegriffo},
  {Mor}, {Morbidelli}, {Morel}, {Morris}, {Mulone}, {Muraveva}, {Musella},
  {Nelemans}, {Nicastro}, {Noval}, {O'Mullane}, {Ord{\'e}novic},
  {Ord{\'o}{\~n}ez-Blanco}, {Osborne}, {Pagani}, {Pagano}, {Pailler},
  {Palacin}, {Palaversa}, {Panahi}, {Pawlak}, {Piersimoni}, {Pineau}, {Plachy},
  {Plum}, {Poggio}, {Poujoulet}, {Pr{\v{s}}a}, {Pulone}, {Racero}, {Ragaini},
  {Rambaux}, {Ramos-Lerate}, {Regibo}, {Reyl{\'e}}, {Riclet}, {Ripepi}, {Riva},
  {Rivard}, {Rixon}, {Roegiers}, {Roelens}, {Romero-G{\'o}mez}, {Rowell},
  {Royer}, {Ruiz-Dern}, {Sadowski}, {Sagrist{\`a} Sell{\'e}s}, {Sahlmann},
  {Salgado}, {Salguero}, {Sanna}, {Santana-Ros}, {Sarasso}, {Savietto},
  {Schultheis}, {Sciacca}, {Segol}, {Segovia}, {S{\'e}gransan}, {Shih},
  {Siltala}, {Silva}, {Smart}, {Smith}, {Solano}, {Solitro}, {Sordo}, {Soria
  Nieto}, {Souchay}, {Spagna}, {Spoto}, {Stampa}, {Steele},
  {Steidelm{\"u}ller}, {Stephenson}, {Stoev}, {Suess}, {Surdej}, {Szabados},
  {Szegedi-Elek}, {Tapiador}, {Taris}, {Tauran}, {Taylor}, {Teixeira},
  {Terrett}, {Teyssandier}, {Thuillot}, {Titarenko}, {Torra Clotet}, {Turon},
  {Ulla}, {Utrilla}, {Uzzi}, {Vaillant}, {Valentini}, {Valette}, {van Elteren},
  {Van Hemelryck}, {van Leeuwen}, {Vaschetto}, {Vecchiato}, {Veljanoski},
  {Viala}, {Vicente}, {Vogt}, {von Essen}, {Voss}, {Votruba}, {Voutsinas},
  {Walmsley}, {Weiler}, {Wertz}, {Wevers}, {Wyrzykowski}, {Yoldas},
  {{\v{Z}}erjal}, {Ziaeepour}, {Zorec}, {Zschocke}, {Zucker}, {Zurbach}, \&
  {Zwitter}}]{2018A&A...616A...1G}
{Gaia Collaboration}, {Brown}, A.~G.~A., {Vallenari}, A., {et~al.} 2018, \aap,
  616, A1

\bibitem[{{Gaia Collaboration} {et~al.}(2021){Gaia Collaboration}, {Smart},
  {Sarro}, {Rybizki}, {Reyl{\'e}}, {Robin}, {Hambly}, {Abbas}, {Barstow}, {de
  Bruijne}, {Bucciarelli}, {Carrasco}, {Cooper}, {Hodgkin}, {Masana},
  {Michalik}, {Sahlmann}, {Sozzetti}, {Brown}, {Vallenari}, {Prusti},
  {Babusiaux}, {Biermann}, {Creevey}, {Evans}, {Eyer}, {Hutton}, {Jansen},
  {Jordi}, {Klioner}, {Lammers}, {Lindegren}, {Luri}, {Mignard}, {Panem},
  {Pourbaix}, {Randich}, {Sartoretti}, {Soubiran}, {Walton}, {Arenou},
  {Bailer-Jones}, {Bastian}, {Cropper}, {Drimmel}, {Katz}, {Lattanzi}, {van
  Leeuwen}, {Bakker}, {Casta{\~n}eda}, {De Angeli}, {Ducourant}, {Fabricius},
  {Fouesneau}, {Fr{\'e}mat}, {Guerra}, {Guerrier}, {Guiraud}, {Jean-Antoine
  Piccolo}, {Messineo}, {Mowlavi}, {Nicolas}, {Nienartowicz}, {Pailler},
  {Panuzzo}, {Riclet}, {Roux}, {Seabroke}, {Sordo}, {Tanga}, {Th{\'e}venin},
  {Gracia-Abril}, {Portell}, {Teyssier}, {Altmann}, {Andrae}, {Bellas-Velidis},
  {Benson}, {Berthier}, {Blomme}, {Brugaletta}, {Burgess}, {Busso}, {Carry},
  {Cellino}, {Cheek}, {Clementini}, {Damerdji}, {Davidson}, {Delchambre},
  {Dell'Oro}, {Fern{\'a}ndez-Hern{\'a}ndez}, {Galluccio}, {Garc{\'\i}a-Lario},
  {Garcia-Reinaldos}, {Gonz{\'a}lez-N{\'u}{\~n}ez}, {Gosset}, {Haigron},
  {Halbwachs}, {Harrison}, {Hatzidimitriou}, {Heiter}, {Hern{\'a}ndez},
  {Hestroffer}, {Holl}, {Jan{\ss}en}, {Jevardat de Fombelle}, {Jordan},
  {Krone-Martins}, {Lanzafame}, {L{\"o}ffler}, {Lorca}, {Manteiga}, {Marchal},
  {Marrese}, {Moitinho}, {Mora}, {Muinonen}, {Osborne}, {Pancino}, {Pauwels},
  {Recio-Blanco}, {Richards}, {Riello}, {Rimoldini}, {Roegiers}, {Siopis},
  {Smith}, {Ulla}, {Utrilla}, {van Leeuwen}, {van Reeven}, {Abreu Aramburu},
  {Accart}, {Aerts}, {Aguado}, {Ajaj}, {Altavilla}, {{\'A}lvarez}, {{\'A}lvarez
  Cid-Fuentes}, {Alves}, {Anderson}, {Anglada Varela}, {Antoja}, {Audard},
  {Baines}, {Baker}, {Balaguer-N{\'u}{\~n}ez}, {Balbinot}, {Balog}, {Barache},
  {Barbato}, {Barros}, {Bartolom{\'e}}, {Bassilana}, {Bauchet},
  {Baudesson-Stella}, {Becciani}, {Bellazzini}, {Bernet}, {Bertone}, {Bianchi},
  {Blanco-Cuaresma}, {Boch}, {Bombrun}, {Bossini}, {Bouquillon}, {Bragaglia},
  {Bramante}, {Breedt}, {Bressan}, {Brouillet}, {Burlacu}, {Busonero},
  {Butkevich}, {Buzzi}, {Caffau}, {Cancelliere}, {C{\'a}novas},
  {Cantat-Gaudin}, {Carballo}, {Carlucci}, {Carnerero}, {Casamiquela},
  {Castellani}, {Castro-Ginard}, {Castro Sampol}, {Chaoul}, {Charlot},
  {Chemin}, {Chiavassa}, {Cioni}, {Comoretto}, {Cornez}, {Cowell}, {Crifo},
  {Crosta}, {Crowley}, {Dafonte}, {Dapergolas}, {David}, {David}, {de Laverny},
  {De Luise}, {De March}, {De Ridder}, {de Souza}, {de Teodoro}, {de Torres},
  {del Peloso}, {del Pozo}, {Delgado}, {Delgado}, {Delisle}, {Di Matteo},
  {Diakite}, {Diener}, {Distefano}, {Dolding}, {Eappachen}, {Edvardsson},
  {Enke}, {Esquej}, {Fabre}, {Fabrizio}, {Faigler}, {Fedorets}, {Fernique},
  {Fienga}, {Figueras}, {Fouron}, {Fragkoudi}, {Fraile}, {Franke}, {Gai},
  {Garabato}, {Garcia-Gutierrez}, {Garc{\'\i}a-Torres}, {Garofalo}, {Gavras},
  {Gerlach}, {Geyer}, {Giacobbe}, {Gilmore}, {Girona}, {Giuffrida}, {Gomel},
  {Gomez}, {Gonzalez-Santamaria}, {Gonz{\'a}lez-Vidal}, {Granvik},
  {Guti{\'e}rrez-S{\'a}nchez}, {Guy}, {Hauser}, {Haywood}, {Helmi}, {Hidalgo},
  {Hilger}, {H{\l}adczuk}, {Hobbs}, {Holland}, {Huckle}, {Jasniewicz},
  {Jonker}, {Juaristi Campillo}, {Julbe}, {Karbevska}, {Kervella}, {Khanna},
  {Kochoska}, {Kontizas}, {Kordopatis}, {Korn}, {Kostrzewa-Rutkowska},
  {Kruszy{\'n}ska}, {Lambert}, {Lanza}, {Lasne}, {Le Campion}, {Le Fustec},
  {Lebreton}, {Lebzelter}, {Leccia}, {Leclerc}, {Lecoeur-Taibi}, {Liao},
  {Licata}, {Lindstr{\o}m}, {Lister}, {Livanou}, {Lobel}, {Madrero Pardo},
  {Managau}, {Mann}, {Marchant}, {Marconi}, {Marcos Santos}, {Marinoni},
  {Marocco}, {Marshall}, {Martin Polo}, {Mart{\'\i}n-Fleitas}, {Masip},
  {Massari}, {Mastrobuono-Battisti}, {Mazeh}, {McMillan}, {Messina}, {Millar},
  {Mints}, {Molina}, {Molinaro}, {Moln{\'a}r}, {Montegriffo}, {Mor},
  {Morbidelli}, {Morel}, {Morris}, {Mulone}, {Munoz}, {Muraveva}, {Murphy},
  {Musella}, {Noval}, {Ord{\'e}novic}, {Orr{\`u}}, {Osinde}, {Pagani},
  {Pagano}, {Palaversa}, {Palicio}, {Panahi}, {Pawlak}, {Pe{\~n}alosa
  Esteller}, {Penttil{\"a}}, {Piersimoni}, {Pineau}, {Plachy}, {Plum},
  {Poggio}, {Poretti}, {Poujoulet}, {Pr{\v{s}}a}, {Pulone}, {Racero},
  {Ragaini}, {Rainer}, {Raiteri}, {Rambaux}, {Ramos}, {Ramos-Lerate}, {Re
  Fiorentin}, {Regibo}, {Ripepi}, {Riva}, {Rixon}, {Robichon}, {Robin},
  {Roelens}, {Rohrbasser}, {Romero-G{\'o}mez}, {Rowell}, {Royer}, {Rybicki},
  {Sadowski}, {Sagrist{\`a} Sell{\'e}s}, {Salgado}, {Salguero}, {Samaras},
  {Sanchez Gimenez}, {Sanna}, {Santove{\~n}a}, {Sarasso}, {Schultheis},
  {Sciacca}, {Segol}, {Segovia}, {S{\'e}gransan}, {Semeux}, {Shahaf},
  {Siddiqui}, {Siebert}, {Siltala}, {Slezak}, {Solano}, {Solitro}, {Souami},
  {Souchay}, {Spagna}, {Spoto}, {Steele}, {Steidelm{\"u}ller}, {Stephenson},
  {S{\"u}veges}, {Szabados}, {Szegedi-Elek}, {Taris}, {Tauran}, {Taylor},
  {Teixeira}, {Thuillot}, {Tonello}, {Torra}, {Torra}, {Turon}, {Unger},
  {Vaillant}, {van Dillen}, {Vanel}, {Vecchiato}, {Viala}, {Vicente},
  {Voutsinas}, {Weiler}, {Wevers}, {Wyrzykowski}, {Yoldas}, {Yvard}, {Zhao},
  {Zorec}, {Zucker}, {Zurbach}, \& {Zwitter}}]{2021A&A...649A...6G}
{Gaia Collaboration}, {Smart}, R.~L., {Sarro}, L.~M., {et~al.} 2021, \aap, 649,
  A6

\bibitem[{{Gaia Collaboration} {et~al.}(2023){Gaia Collaboration}, Vallenari,
  Brown, Prusti, de~Bruijne, Arenou, Babusiaux, Biermann, Creevey, Ducourant,
  Evans, Eyer, Guerra, Hutton, Jordi, Klioner, Lammers, Lindegren, Luri,
  Mignard, Panem, Pourbaix, Randich, Sartoretti, Soubiran, Tanga, Walton,
  Bailer-Jones, Bastian, Drimmel, Jansen, Katz, Lattanzi, van Leeuwen, Bakker,
  Cacciari, Casta{\~n}eda, De~Angeli, Fabricius, Fouesneau, Fr{\'e}mat,
  Galluccio, Guerrier, Heiter, Masana, Messineo, Mowlavi, Nicolas,
  Nienartowicz, Pailler, Panuzzo, Riclet, Roux, Seabroke, Sordo, Th{\'e}venin,
  Gracia-Abril, Portell, Teyssier, Altmann, Andrae, Audard, Bellas-Velidis,
  Benson, Berthier, Blomme, Burgess, Busonero, Busso, C{\'a}novas, Carry,
  Cellino, Cheek, Clementini, Damerdji, Davidson, de~Teodoro, Nu{\~n}ez~Campos,
  Delchambre, Dell'Oro, Esquej, Fern{\'a}ndez-Hern{\'a}ndez, Fraile, Garabato,
  Garc{\'\i}a-Lario, Gosset, Haigron, Halbwachs, Hambly, Harrison,
  Hern{\'a}ndez, Hestroffer, Hodgkin, Holl, Jan{\ss}en, Jevardat~de Fombelle,
  Jordan, Krone-Martins, Lanzafame, L{\"o}ffler, Marchal, Marrese, Moitinho,
  Muinonen, Osborne, Pancino, Pauwels, Recio-Blanco, Reyl{\'e}, Riello,
  Rimoldini, Roegiers, Rybizki, Sarro, Siopis, Smith, Sozzetti, Utrilla, van
  Leeuwen, Abbas, {\'A}brah{\'a}m, Abreu~Aramburu, Aerts, Aguado, Ajaj,
  Aldea-Montero, Altavilla, {\'A}lvarez, Alves, Anders, Anderson,
  Anglada~Varela, Antoja, Baines, Baker, Balaguer-N{\'u}{\~n}ez, Balbinot,
  Balog, Barache, Barbato, Barros, Barstow, Bartolom{\'e}, Bassilana, Bauchet,
  Becciani, Bellazzini, Berihuete, Bernet, Bertone, Bianchi, Binnenfeld,
  Blanco-Cuaresma, Blazere, Boch, Bombrun, Bossini, Bouquillon, Bragaglia,
  Bramante, Breedt, Bressan, Brouillet, Brugaletta, Bucciarelli, Burlacu,
  Butkevich, Buzzi, Caffau, Cancelliere, Cantat-Gaudin, Carballo, Carlucci,
  Carnerero, Carrasco, Casamiquela, Castellani, Castro-Ginard, Chaoul, Charlot,
  Chemin, Chiaramida, Chiavassa, Chornay, Comoretto, Contursi, Cooper, Cornez,
  Cowell, Crifo, Cropper, Crosta, Crowley, Dafonte, Dapergolas, David, David,
  de~Laverny, De~Luise, De~March, De~Ridder, de~Souza, de~Torres, del Peloso,
  del Pozo, Delbo, Delgado, Delisle, Demouchy, Dharmawardena, Di~Matteo,
  Diakite, Diener, Distefano, Dolding, Edvardsson, Enke, Fabre, Fabrizio,
  Faigler, Fedorets, Fernique, Fienga, Figueras, Fournier, Fouron, Fragkoudi,
  Gai, Garcia-Gutierrez, Garcia-Reinaldos, Garc{\'\i}a-Torres, Garofalo, Gavel,
  Gavras, Gerlach, Geyer, Giacobbe, Gilmore, Girona, Giuffrida, Gomel, Gomez,
  Gonz{\'a}lez-N{\'u}{\~n}ez, Gonz{\'a}lez-Santamar{\'\i}a, Gonz{\'a}lez-Vidal,
  Granvik, Guillout, Guiraud, Guti{\'e}rrez-S{\'a}nchez, Guy, Hatzidimitriou,
  Hauser, Haywood, Helmer, Helmi, Sarmiento, Hidalgo, Hilger, H{\l}adczuk,
  Hobbs, Holland, Huckle, Jardine, Jasniewicz, Jean-Antoine~Piccolo,
  Jim{\'e}nez-Arranz, Jorissen, Juaristi~Campillo, Julbe, Karbevska, Kervella,
  Khanna, Kontizas, Kordopatis, Korn, K{\'o}sp{\'a}l, Kostrzewa-Rutkowska,
  Kruszy{\'n}ska, Kun, Laizeau, Lambert, Lanza, Lasne, Le~Campion, Lebreton,
  Lebzelter, Leccia, Leclerc, Lecoeur-Taibi, Liao, Licata, Lindstr{\o}m,
  Lister, Livanou, Lobel, Lorca, Loup, Madrero~Pardo, Magdaleno~Romeo, Managau,
  Mann, Manteiga, Marchant, Marconi, Marcos, Marcos~Santos, Mar{\'\i}n~Pina,
  Marinoni, Marocco, Marshall, Martin~Polo, Mart{\'\i}n-Fleitas, Marton, Mary,
  Masip, Massari, Mastrobuono-Battisti, Mazeh, McMillan, Messina, Michalik,
  Millar, Mints, Molina, Molinaro, Moln{\'a}r, Monari, Mongui{\'o},
  Montegriffo, Montero, Mor, Mora, Morbidelli, Morel, Morris, Muraveva, Murphy,
  Musella, Nagy, Noval, Oca{\~n}a, Ogden, Ordenovic, Osinde, Pagani, Pagano,
  Palaversa, Palicio, Pallas-Quintela, Panahi, Payne-Wardenaar,
  Pe{\~n}alosa~Esteller, Penttil{\"a}, Pichon, Piersimoni, Pineau, Plachy,
  Plum, Poggio, Pr{\v{s}}a, Pulone, Racero, Ragaini, Rainer, Raiteri, Rambaux,
  Ramos, Ramos-Lerate, Re~Fiorentin, Regibo, Richards, Rios~Diaz, Ripepi, Riva,
  Rix, Rixon, Robichon, Robin, Robin, Roelens, Rogues, Rohrbasser,
  Romero-G{\'o}mez, Rowell, Royer, Ruz~Mieres, Rybicki, Sadowski,
  S{\'a}ez~N{\'u}{\~n}ez, Sagrist{\`a}~Sell{\'e}s, Sahlmann, Salguero, Samaras,
  Sanchez~Gimenez, Sanna, Santove{\~n}a, Sarasso, Schultheis, Sciacca, Segol,
  Segovia, S{\'e}gransan, Semeux, Shahaf, Siddiqui, Siebert, Siltala, Silvelo,
  Slezak, Slezak, Smart, Snaith, Solano, Solitro, Souami, Souchay, Spagna,
  Spina, Spoto, Steele, Steidelm{\"u}ller, Stephenson, S{\"u}veges, Surdej,
  Szabados, Szegedi-Elek, Taris, Taylor, Teixeira, Tolomei, Tonello, Torra,
  Torra, Torralba~Elipe, Trabucchi, Tsounis, Turon, Ulla, Unger, Vaillant, van
  Dillen, van Reeven, Vanel, Vecchiato, Viala, Vicente, Voutsinas, Weiler,
  Wevers, Wyrzykowski, Yoldas, Yvard, Zhao, Zorec, Zucker, \&
  Zwitter}]{GaiaCollaboration2023}
{Gaia Collaboration}, Vallenari, A., Brown, A.~G.~A., {et~al.} 2023, \aap, 674,
  A1

\bibitem[{García {et~al.}(2012)García, Sánchez, \& Mollineda}]{GARCIA201213}
García, V., Sánchez, J., \& Mollineda, R. 2012, Knowledge-Based Systems, 25,
  13, special Issue on New Trends in Data Mining

\bibitem[{{Gonzalez} {et~al.}(2022){Gonzalez}, {Cirasuolo}, {Taylor}, {Black},
  {Rees}, {Bryson}, {Chittick}, {Afonso}, {Lilly}, {Flores}, {Maiolino},
  {Oliva}, {Paltani}, {Vanzi}, {Abreu}, {Amans}, {Atkinson}, {Beard},
  {Belfiore}, {Breen}, {Bayo}, {Born}, {Cabral}, {Chapman}, {Cochrane},
  {Coelho}, {Colling}, {Conzelmann}, {Dalessio}, {Davidson},
  {Delplancke-Str{\"o}bele}, {Fisher}, {Forchi}, {Franzetti}, {Garilli},
  {Gargiulo}, {Guinouard}, {Gutierrez}, {Haigron}, {Hammersley}, {Ivanov},
  {Ives}, {Iwert}, {King}, {Kovacz}, {Laporte}, {Lee}, {Li Causi}, {Macleod},
  {Alvarez Mendez}, {Oliveira}, {Palsa}, {Parra}, {Pedichini}, {Pe{\~n}a},
  {Petr-Gotzens}, {Rodrigues}, {Royer}, {Santos}, {Sepulveda}, {Sharman},
  {Shen}, {Sordet}, {Strachan}, {Tait}, {Tejeda}, {Tozzi}, {O'Malley},
  {Waring}, {Watson}, {Willemse}, {Gao}, {Yang}, \& {Zoccali}}]{moons2022}
{Gonzalez}, O., {Cirasuolo}, M., {Taylor}, W., {et~al.} 2022, in Society of
  Photo-Optical Instrumentation Engineers (SPIE) Conference Series, Vol. 12184,
  Ground-based and Airborne Instrumentation for Astronomy IX, ed. C.~J.
  {Evans}, J.~J. {Bryant}, \& K.~{Motohara}, 1218412

\bibitem[{{Gray} \& {Corbally}(2009)}]{2009ssc..book.....G}
{Gray}, R.~O. \& {Corbally}, Christopher, J. 2009, {Stellar Spectral
  Classification}

\bibitem[{Harris {et~al.}(2020)Harris, Millman, van~der Walt, Gommers,
  Virtanen, Cournapeau, Wieser, Taylor, Berg, Smith, Kern, Picus, Hoyer, van
  Kerkwijk, Brett, Haldane, del R{\'{i}}o, Wiebe, Peterson,
  G{\'{e}}rard-Marchant, Sheppard, Reddy, Weckesser, Abbasi, Gohlke, \&
  Oliphant}]{harris2020array}
Harris, C.~R., Millman, K.~J., van~der Walt, S.~J., {et~al.} 2020, Nature, 585,
  357

\bibitem[{Heger {et~al.}(2003)Heger, Fryer, Woosley, Langer, \&
  Hartmann}]{Heger_2003}
Heger, A., Fryer, C.~L., Woosley, S.~E., Langer, N., \& Hartmann, D.~H. 2003,
  The Astrophysical Journal, 591, 288

\bibitem[{Hunter(2007)}]{Hunter:2007}
Hunter, J.~D. 2007, Computing in Science \& Engineering, 9, 90

\bibitem[{Izzard {et~al.}(2004)Izzard, Ramirez-Ruiz, \& Tout}]{Izzard2004}
Izzard, R.~G., Ramirez-Ruiz, E., \& Tout, C.~A. 2004, Monthly Notices of the
  Royal Astronomical Society, 348, 1215–1228

\bibitem[{{Jin} {et~al.}(2024){Jin}, {Trager}, {Dalton}, {Aguerri}, {Drew},
  {Falc{\'o}n-Barroso}, {G{\"a}nsicke}, {Hill}, {Iovino}, {Pieri}, {Poggianti},
  {Smith}, {Vallenari}, {Abrams}, {Aguado}, {Antoja}, {Arag{\'o}n-Salamanca},
  {Ascasibar}, {Babusiaux}, {Balcells}, {Barrena}, {Battaglia}, {Belokurov},
  {Bensby}, {Bonifacio}, {Bragaglia}, {Carrasco}, {Carrera}, {Cornwell},
  {Dom{\'\i}nguez-Palmero}, {Duncan}, {Famaey}, {Fari{\~n}a}, {Gonzalez},
  {Guest}, {Hatch}, {Hess}, {Hoskin}, {Irwin}, {Knapen}, {Koposov}, {Kuchner},
  {Laigle}, {Lewis}, {Longhetti}, {Lucatello}, {M{\'e}ndez-Abreu}, {Mercurio},
  {Molaeinezhad}, {Mongui{\'o}}, {Morrison}, {Murphy}, {Peralta de Arriba},
  {P{\'e}rez}, {P{\'e}rez-R{\`a}fols}, {Pic{\'o}}, {Raddi}, {Romero-G{\'o}mez},
  {Royer}, {Siebert}, {Seabroke}, {Som}, {Terrett}, {Thomas}, {Wesson},
  {Worley}, {Alfaro}, {Allende Prieto}, {Alonso-Santiago}, {Amos}, {Ashley},
  {Balaguer-N{\'u}{\~n}ez}, {Balbinot}, {Bellazzini}, {Benn}, {Berlanas},
  {Bernard}, {Best}, {Bettoni}, {Bianco}, {Bishop}, {Blomqvist}, {Boeche},
  {Bolzonella}, {Bonoli}, {Bosma}, {Britavskiy}, {Busarello}, {Caffau},
  {Cantat-Gaudin}, {Castro-Ginard}, {Couto}, {Carbajo-Hijarrubia}, {Carter},
  {Casamiquela}, {Conrado}, {Corcho-Caballero}, {Costantin}, {Deason}, {de
  Burgos}, {De Grandi}, {Di Matteo}, {Dom{\'\i}nguez-G{\'o}mez}, {Dorda},
  {Drake}, {Dutta}, {Erkal}, {Feltzing}, {Ferr{\'e}-Mateu}, {Feuillet},
  {Figueras}, {Fossati}, {Franciosini}, {Frasca}, {Fumagalli}, {Gallazzi},
  {Garc{\'\i}a-Benito}, {Gentile Fusillo}, {Gebran}, {Gilbert}, {Gledhill},
  {Gonz{\'a}lez Delgado}, {Greimel}, {Guarcello}, {Guerra}, {Gullieuszik},
  {Haines}, {Hardcastle}, {Harris}, {Haywood}, {Helmi}, {Hernandez}, {Herrero},
  {Hughes}, {Ir{\v{s}}i{\v{c}}}, {Jablonka}, {Jarvis}, {Jordi}, {Kondapally},
  {Kordopatis}, {Krogager}, {La Barbera}, {Lam}, {Larsen}, {Lemasle}, {Lewis},
  {Lhom{\'e}}, {Lind}, {Lodi}, {Longobardi}, {Lonoce}, {Magrini}, {Ma{\'\i}z
  Apell{\'a}niz}, {Marchal}, {Marco}, {Martin}, {Matsuno}, {Maurogordato},
  {Merluzzi}, {Miralda-Escud{\'e}}, {Molinari}, {Monari}, {Morelli}, {Mottram},
  {Naylor}, {Negueruela}, {O{\~n}orbe}, {Pancino}, {Peirani}, {Peletier},
  {Pozzetti}, {Rainer}, {Ramos}, {Read}, {Rossi}, {R{\"o}ttgering},
  {Rubi{\~n}o-Mart{\'\i}n}, {Sabater}, {San Juan}, {Sanna}, {Schallig},
  {Schiavon}, {Schultheis}, {Serra}, {Shimwell}, {Sim{\'o}n-D{\'\i}az},
  {Smith}, {Sordo}, {Sorini}, {Soubiran}, {Starkenburg}, {Steele}, {Stott},
  {Stuik}, {Tolstoy}, {Tortora}, {Tsantaki}, {Van der Swaelmen}, {van Weeren},
  {Vergani}, {Verheijen}, {Verro}, {Vink}, {Vioque}, {Walcher}, {Walton},
  {Wegg}, {Weijmans}, {Williams}, {Wilson}, {Wright}, {Xylakis-Dornbusch},
  {Youakim}, {Zibetti}, \& {Zurita}}]{weave2024}
{Jin}, S., {Trager}, S.~C., {Dalton}, G.~B., {et~al.} 2024, \mnras, 530, 2688

\bibitem[{{Kaufer} {et~al.}(1999){Kaufer}, {Stahl}, {Tubbesing},
  {N{\o}rregaard}, {Avila}, {Francois}, {Pasquini}, \&
  {Pizzella}}]{1999Msngr..95....8K}
{Kaufer}, A., {Stahl}, O., {Tubbesing}, S., {et~al.} 1999, The Messenger, 95, 8

\bibitem[{{Kollmeier} {et~al.}(2019){Kollmeier}, {Anderson}, {Blanc},
  {Blanton}, {Covey}, {Crane}, {Drory}, {Frinchaboy}, {Froning}, {Johnson},
  {Kneib}, {Kreckel}, {Merloni}, {Pellegrini}, {Pogge}, {Ramirez}, {Rix},
  {Sayres}, {S{\'a}nchez-Gallego}, {Shen}, {Tkachenko}, {Trump}, {Tuttle},
  {Weijmans}, {Zasowski}, {Barbuy}, {Beaton}, {Bergemann}, {Bochanski},
  {Brandt}, {Casey}, {Cherinka}, {Eracleous}, {Fan}, {Garc{\'\i}a}, {Green},
  {Hekker}, {Lane}, {Longa-Pe{\~n}a}, {Mathur}, {Meza}, {Minchev}, {Myers},
  {Nidever}, {Nitschelm}, {O'Connell}, {Price-Whelan}, {Raddick}, {Rossi},
  {Sankrit}, {Simon}, {Stutz}, {Ting}, {Trakhtenbrot}, {Weaver}, {Willmer}, \&
  {Weinberg}}]{SSDSV2019}
{Kollmeier}, J., {Anderson}, S.~F., {Blanc}, G.~A., {et~al.} 2019, in Bulletin
  of the American Astronomical Society, Vol.~51, 274

\bibitem[{{Kounkel} {et~al.}(2023){Kounkel}, {Zari}, {Covey}, {Tkachenko},
  {Z{\'u}{\~n}iga}, {Stassun}, {Stutz}, {Stringfellow}, {Roman-Lopes},
  {Hern{\'a}ndez}, {Pe{\~n}a Ram{\'\i}rez}, {Bayo}, {Kim}, {Cao}, {Wolk},
  {Kollmeier}, {L{\'o}pez-Valdivia}, \& {Rojas-Ayala}}]{2023ApJS..266...10K}
{Kounkel}, M., {Zari}, E., {Covey}, K., {et~al.} 2023, \apjs, 266, 10

\bibitem[{{Kroupa} {et~al.}(2024){Kroupa}, {Gjergo}, {Jerabkova}, \&
  {Yan}}]{Kroupa2024}
{Kroupa}, P., {Gjergo}, E., {Jerabkova}, T., \& {Yan}, Z. 2024, arXiv e-prints,
  arXiv:2410.07311

\bibitem[{{Kroupa} \& {Weidner}(2003)}]{Kroupa2003}
{Kroupa}, P. \& {Weidner}, C. 2003, \apj, 598, 1076

\bibitem[{{Kyritsis} {et~al.}(2022){Kyritsis}, {Maravelias}, {Zezas},
  {Bonfini}, {Kovlakas}, \& {Reig}}]{2022A&A...657A..62K}
{Kyritsis}, E., {Maravelias}, G., {Zezas}, A., {et~al.} 2022, \aap, 657, A62

\bibitem[{Lema{{\^i}}tre {et~al.}(2017)Lema{{\^i}}tre, Nogueira, \&
  Aridas}]{JMLR:v18:16-365}
Lema{{\^i}}tre, G., Nogueira, F., \& Aridas, C.~K. 2017, Journal of Machine
  Learning Research, 18, 1

\bibitem[{Li(2021)}]{Li_2021}
Li, G.-W. 2021, The Astrophysical Journal Supplement Series, 253, 54

\bibitem[{{Lucas} {et~al.}(2024){Lucas}, {Smith}, {Guo}, {Contreras Pe{\~n}a},
  {Minniti}, {Miller}, {Alonso-Garc{\'\i}a}, {Catelan}, {Borissova}, {Saito},
  {Kurtev}, {Navarro}, {Morris}, {Muthu}, {Froebrich}, {Ivanov}, {Bayo},
  {Caratti o Garatti}, \& {Sanders}}]{2024MNRAS.528.1789L}
{Lucas}, P.~W., {Smith}, L.~C., {Guo}, Z., {et~al.} 2024, \mnras, 528, 1789

\bibitem[{{Ma{\'\i}z Apell{\'a}niz} {et~al.}(2024){Ma{\'\i}z Apell{\'a}niz},
  {Negueruela}, \& {Caballero}}]{Jesus2024}
{Ma{\'\i}z Apell{\'a}niz}, J., {Negueruela}, I., \& {Caballero}, J.~A. 2024,
  arXiv e-prints, arXiv:2410.07301

\bibitem[{{Ma{\'\i}z Apell{\'a}niz} {et~al.}(2016){Ma{\'\i}z Apell{\'a}niz},
  {Sota}, {Arias}, {Barb{\'a}}, {Walborn}, {Sim{\'o}n-D{\'\i}az}, {Negueruela},
  {Marco}, {Le{\~a}o}, {Herrero}, {Gamen}, \& {Alfaro}}]{2016ApJS..224....4M}
{Ma{\'\i}z Apell{\'a}niz}, J., {Sota}, A., {Arias}, J.~I., {et~al.} 2016,
  \apjs, 224, 4

\bibitem[{{Ma{\'\i}z Apell{\'a}niz} {et~al.}(2013){Ma{\'\i}z Apell{\'a}niz},
  {Sota}, {Morrell}, {Barb{\'a}}, {Walborn}, {Alfaro}, {Gamen}, {Arias}, \&
  {Gallego Calvente}}]{Goss2013}
{Ma{\'\i}z Apell{\'a}niz}, J., {Sota}, A., {Morrell}, N.~I., {et~al.} 2013, in
  Massive Stars: From alpha to Omega, 198

\bibitem[{{Martins}(2018)}]{2018A&A...616A.135M}
{Martins}, F. 2018, \aap, 616, A135

\bibitem[{{Minniti} {et~al.}(2010){Minniti}, {Lucas}, {Emerson}, {Saito},
  {Hempel}, {Pietrukowicz}, {Ahumada}, {Alonso}, {Alonso-Garcia}, {Arias},
  {Bandyopadhyay}, {Barb{\'a}}, {Barbuy}, {Bedin}, {Bica}, {Borissova},
  {Bronfman}, {Carraro}, {Catelan}, {Clari{\'a}}, {Cross}, {de Grijs},
  {D{\'e}k{\'a}ny}, {Drew}, {Fari{\~n}a}, {Feinstein}, {Fern{\'a}ndez
  Laj{\'u}s}, {Gamen}, {Geisler}, {Gieren}, {Goldman}, {Gonzalez}, {Gunthardt},
  {Gurovich}, {Hambly}, {Irwin}, {Ivanov}, {Jord{\'a}n}, {Kerins}, {Kinemuchi},
  {Kurtev}, {L{\'o}pez-Corredoira}, {Maccarone}, {Masetti}, {Merlo},
  {Messineo}, {Mirabel}, {Monaco}, {Morelli}, {Padilla}, {Palma}, {Parisi},
  {Pignata}, {Rejkuba}, {Roman-Lopes}, {Sale}, {Schreiber}, {Schr{\"o}der},
  {Smith}, {Sodr{\'e}}, {Soto}, {Tamura}, {Tappert}, {Thompson}, {Toledo},
  {Zoccali}, \& {Pietrzynski}}]{2010NewA...15..433M}
{Minniti}, D., {Lucas}, P.~W., {Emerson}, J.~P., {et~al.} 2010, \na, 15, 433

\bibitem[{Moe \& {Di Stefano}(2017)}]{Moe2017}
Moe, M. \& {Di Stefano}, R. 2017, \apjs, 230, 15

\bibitem[{{Morgan}(1951)}]{morgan51}
{Morgan}, W.~W. 1951, Publications of Michigan Observatory, 10, 33

\bibitem[{{Morgan} \& {Keenan}(1973)}]{1973ARA&A..11...29M}
{Morgan}, W.~W. \& {Keenan}, P.~C. 1973, \araa, 11, 29

\bibitem[{{Moritz} {et~al.}(2017){Moritz}, {Nishihara}, {Wang}, {Tumanov},
  {Liaw}, {Liang}, {Elibol}, {Yang}, {Paul}, {Jordan}, \&
  {Stoica}}]{2017arXiv171205889M}
{Moritz}, P., {Nishihara}, R., {Wang}, S., {et~al.} 2017, arXiv e-prints,
  arXiv:1712.05889

\bibitem[{{Nakar}(2007)}]{Nakar2007}
{Nakar}, E. 2007, \physrep, 442, 166–236

\bibitem[{{Negueruela} {et~al.}(2024){Negueruela}, {Sim{\'o}n-D{\'\i}az}, {de
  Burgos}, {Casasbuenas}, \& {Beck}}]{2024A&A...690A.176N}
{Negueruela}, I., {Sim{\'o}n-D{\'\i}az}, S., {de Burgos}, A., {Casasbuenas},
  A., \& {Beck}, P.~G. 2024, \aap, 690, A176

\bibitem[{{Offner} {et~al.}(2023){Offner}, {Moe}, {Kratter}, {Sadavoy},
  {Jensen}, \& {Tobin}}]{Offner2023}
{Offner}, S.~S.~R., {Moe}, M., {Kratter}, K.~M., {et~al.} 2023, in Astronomical
  Society of the Pacific Conference Series, Vol. 534, Protostars and Planets
  VII, ed. S.~{Inutsuka}, Y.~{Aikawa}, T.~{Muto}, K.~{Tomida}, \& M.~{Tamura},
  275

\bibitem[{pandas~development team(2020)}]{reback2020pandas}
pandas~development team, T. 2020, pandas-dev/pandas: Pandas

\bibitem[{{Pantaleoni Gonz{\'a}lez} {et~al.}(2021){Pantaleoni Gonz{\'a}lez},
  {Ma{\'\i}z Apell{\'a}niz}, {Barb{\'a}}, \& {Reed}}]{2021MNRAS.504.2968P}
{Pantaleoni Gonz{\'a}lez}, M., {Ma{\'\i}z Apell{\'a}niz}, J., {Barb{\'a}},
  R.~H., \& {Reed}, B.~C. 2021, \mnras, 504, 2968

\bibitem[{Pedregosa {et~al.}(2011)Pedregosa, Varoquaux, Gramfort, Michel,
  Thirion, Grisel, Blondel, Prettenhofer, Weiss, Dubourg, Vanderplas, Passos,
  Cournapeau, Brucher, Perrot, \& Duchesnay}]{scikit-learn}
Pedregosa, F., Varoquaux, G., Gramfort, A., {et~al.} 2011, Journal of Machine
  Learning Research, 12, 2825

\bibitem[{{Pickering}(1890)}]{1890AnHar..27....1P}
{Pickering}, E.~C. 1890, Annals of Harvard College Observatory, 27, 1

\bibitem[{{Radley} {et~al.}(2025){Radley}, {Oudmaijer}, {Vioque}, \&
  {Dodd}}]{2025MNRAS.539.1964R}
{Radley}, I.~C., {Oudmaijer}, R.~D., {Vioque}, M., \& {Dodd}, J.~M. 2025,
  \mnras, 539, 1964

\bibitem[{{Ram{\'\i}rez-Preciado} {et~al.}(2020){Ram{\'\i}rez-Preciado},
  {Roman-Lopes}, {Rom{\'a}n-Z{\'u}{\~n}iga}, {Hern{\'a}ndez},
  {Garc{\'\i}a-Hern{\'a}ndez}, {Stassun}, {Stringfellow}, \&
  {Kim}}]{2020ApJ...894....5R}
{Ram{\'\i}rez-Preciado}, V.~G., {Roman-Lopes}, A., {Rom{\'a}n-Z{\'u}{\~n}iga},
  C.~G., {et~al.} 2020, \apj, 894, 5

\bibitem[{{Raskin} {et~al.}(2011){Raskin}, {van Winckel}, {Hensberge},
  {Jorissen}, {Lehmann}, {Waelkens}, {Avila}, {de Cuyper}, {Degroote},
  {Dubosson}, {Dumortier}, {Fr{\'e}mat}, {Laux}, {Michaud}, {Morren}, {Perez
  Padilla}, {Pessemier}, {Prins}, {Smolders}, {van Eck}, \&
  {Winkler}}]{2011A&A...526A..69R}
{Raskin}, G., {van Winckel}, H., {Hensberge}, H., {et~al.} 2011, \aap, 526, A69

\bibitem[{{Roman-Lopes} \& {Roman-Lopes}(2019)}]{2019MNRAS.484.5578R}
{Roman-Lopes}, A. \& {Roman-Lopes}, G.~F. 2019, \mnras, 484, 5578

\bibitem[{{Rybizki} \& {Just}(2015)}]{Rybizki2015}
{Rybizki}, J. \& {Just}, A. 2015, \mnras, 447, 3880

\bibitem[{Sana {et~al.}(2012)Sana, de~Mink, de~Koter, Langer, Evans, Gieles,
  Gosset, Izzard, {Le Bouquin}, \& Schneider}]{Sana2012}
Sana, H., de~Mink, S.~E., de~Koter, A., {et~al.} 2012, Science, 337, 444

\bibitem[{{S{\'a}nchez-S{\'a}ez} {et~al.}(2021){S{\'a}nchez-S{\'a}ez}, {Reyes},
  {Valenzuela}, {F{\"o}rster}, {Eyheramendy}, {Elorrieta}, {Bauer},
  {Cabrera-Vives}, {Est{\'e}vez}, {Catelan}, {Pignata}, {Huijse}, {De Cicco},
  {Ar{\'e}valo}, {Carrasco-Davis}, {Abril}, {Kurtev}, {Borissova}, {Arredondo},
  {Castillo-Navarrete}, {Rodriguez}, {Ruz-Mieres}, {Moya},
  {Sabatini-Gacit{\'u}a}, {Sep{\'u}lveda-Cobo}, \&
  {Camacho-I{\~n}iguez}}]{2021AJ....161..141S}
{S{\'a}nchez-S{\'a}ez}, P., {Reyes}, I., {Valenzuela}, C., {et~al.} 2021, \aj,
  161, 141

\bibitem[{Schneider {et~al.}(2001)Schneider, Ferrari, Matarrese, \& {Portegies
  Zwart}}]{Schneider2001}
Schneider, R., Ferrari, V., Matarrese, S., \& {Portegies Zwart}, S.~F. 2001,
  \mnras, 324, 797–810

\bibitem[{{Sim{\'o}n-D{\'\i}az}
  {et~al.}(2011{\natexlab{a}}){Sim{\'o}n-D{\'\i}az}, {Castro}, {Garcia},
  {Herrero}, \& {Markova}}]{IACOB2011}
{Sim{\'o}n-D{\'\i}az}, S., {Castro}, N., {Garcia}, M., {Herrero}, A., \&
  {Markova}, N. 2011{\natexlab{a}}, Bulletin de la Societe Royale des Sciences
  de Liege, 80, 514

\bibitem[{{Sim{\'o}n-D{\'\i}az}
  {et~al.}(2011{\natexlab{b}}){Sim{\'o}n-D{\'\i}az}, {Garcia}, {Herrero},
  {Ma{\'\i}z Apell{\'a}niz}, \& {Negueruela}}]{2011sca..conf..255S}
{Sim{\'o}n-D{\'\i}az}, S., {Garcia}, M., {Herrero}, A., {Ma{\'\i}z
  Apell{\'a}niz}, J., \& {Negueruela}, I. 2011{\natexlab{b}}, in Stellar
  Clusters \& Associations: A RIA Workshop on Gaia, 255--259

\bibitem[{{Sim{\'o}n-D{\'\i}az} {et~al.}(2015){Sim{\'o}n-D{\'\i}az},
  {Negueruela}, {Ma{\'\i}z Apell{\'a}niz}, {Castro}, {Herrero}, {Garcia},
  {P{\'e}rez-Prieto}, {Caon}, {Alacid}, {Camacho}, {Dorda}, {Godart},
  {Gonz{\'a}lez-Fern{\'a}ndez}, {Holgado}, \&
  {R{\"u}bke}}]{2015hsa8.conf..576S}
{Sim{\'o}n-D{\'\i}az}, S., {Negueruela}, I., {Ma{\'\i}z Apell{\'a}niz}, J.,
  {et~al.} 2015, in Highlights of Spanish Astrophysics VIII, ed. A.~J.
  {Cenarro}, F.~{Figueras}, C.~{Hern{\'a}ndez-Monteagudo}, J.~{Trujillo Bueno},
  \& L.~{Valdivielso}, 576--581

\bibitem[{{Skiff}(2014)}]{2014yCat....1.2023S}
{Skiff}, B.~A. 2014, VizieR Online Data Catalog, B/mk

\bibitem[{{Skrutskie} {et~al.}(2006){Skrutskie}, {Cutri}, {Stiening},
  {Weinberg}, {Schneider}, {Carpenter}, {Beichman}, {Capps}, {Chester},
  {Elias}, {Huchra}, {Liebert}, {Lonsdale}, {Monet}, {Price}, {Seitzer},
  {Jarrett}, {Kirkpatrick}, {Gizis}, {Howard}, {Evans}, {Fowler}, {Fullmer},
  {Hurt}, {Light}, {Kopan}, {Marsh}, {McCallon}, {Tam}, {Van Dyk}, \&
  {Wheelock}}]{2006AJ....131.1163S}
{Skrutskie}, M.~F., {Cutri}, R.~M., {Stiening}, R., {et~al.} 2006, \aj, 131,
  1163

\bibitem[{{Smartt}(2009)}]{2009ARA&A..47...63S}
{Smartt}, S.~J. 2009, \araa, 47, 63

\bibitem[{{Sota} {et~al.}(2011){Sota}, {Ma{\'\i}z Apell{\'a}niz}, {Walborn},
  {Alfaro}, {Barb{\'a}}, {Morrell}, {Gamen}, \& {Arias}}]{sota2011}
{Sota}, A., {Ma{\'\i}z Apell{\'a}niz}, J., {Walborn}, N.~R., {et~al.} 2011,
  \apjs, 193, 24

\bibitem[{{Sota} {et~al.}(2014){Sota}, {Maíz Apellániz}, {Morrell}, {Barbá},
  {Walborn}, {Gamen}, {Arias}, \& {Alfaro}}]{sota2014}
{Sota}, A., {Maíz Apellániz}, J., {Morrell}, N.~I., {et~al.} 2014, \apjs,
  211, 10

\bibitem[{{Taylor}(2005)}]{2005ASPC..347...29T}
{Taylor}, M.~B. 2005, in Astronomical Society of the Pacific Conference Series,
  Vol. 347, Astronomical Data Analysis Software and Systems XIV, ed.
  P.~{Shopbell}, M.~{Britton}, \& R.~{Ebert}, 29

\bibitem[{Virtanen {et~al.}(2020)Virtanen, Gommers, Oliphant, Haberland, Reddy,
  Cournapeau, Burovski, Peterson, Weckesser, Bright, {van der Walt}, Brett,
  Wilson, Millman, Mayorov, Nelson, Jones, Kern, Larson, Carey, Polat, Feng,
  Moore, {VanderPlas}, Laxalde, Perktold, Cimrman, Henriksen, Quintero, Harris,
  Archibald, Ribeiro, Pedregosa, {van Mulbregt}, \& {SciPy 1.0
  Contributors}}]{2020SciPy-NMeth}
Virtanen, P., Gommers, R., Oliphant, T.~E., {et~al.} 2020, Nature Methods, 17,
  261

\bibitem[{{Walborn}(1972)}]{Walborn1972}
{Walborn}, N.~R. 1972, \aj, 77, 312

\bibitem[{{Wenger} {et~al.}(2000){Wenger}, {Ochsenbein}, {Egret}, {Dubois},
  {Bonnarel}, {Borde}, {Genova}, {Jasniewicz}, {Lalo{\"e}}, {Lesteven}, \&
  {Monier}}]{2000A&AS..143....9W}
{Wenger}, M., {Ochsenbein}, F., {Egret}, D., {et~al.} 2000, \aaps, 143, 9

\bibitem[{{Zari} {et~al.}(2021){Zari}, {Rix}, {Frankel}, {Xiang}, {Poggio},
  {Drimmel}, \& {Tkachenko}}]{zari2021}
{Zari}, E., {Rix}, H.~W., {Frankel}, N., {et~al.} 2021, \aap, 650, A112

\bibitem[{{Zhao} {et~al.}(2012){Zhao}, {Zhao}, {Chu}, {Jing}, \&
  {Deng}}]{2012RAA....12..723Z}
{Zhao}, G., {Zhao}, Y.-H., {Chu}, Y.-Q., {Jing}, Y.-P., \& {Deng}, L.-C. 2012,
  Research in Astronomy and Astrophysics, 12, 723

\bibitem[{{Zhu} {et~al.}(2006){Zhu}, {Hu}, {Zhang}, {Wang}, \&
  {Wang}}]{2006SPIE.6269E..0MZ}
{Zhu}, Y., {Hu}, Z., {Zhang}, Q., {Wang}, L., \& {Wang}, J. 2006, in Society of
  Photo-Optical Instrumentation Engineers (SPIE) Conference Series, Vol. 6269,
  Ground-based and Airborne Instrumentation for Astronomy, ed. I.~S. {McLean}
  \& M.~{Iye}, 62690M

\bibitem[{{Zinnecker} \& {Yorke}(2007)}]{Zinnecker2007}
{Zinnecker}, H. \& {Yorke}, H.~W. 2007, \araa, 45, 481–563

\end{thebibliography}
\appendix
\onecolumn
\section{Complementary spectroscopic codification} \label{apendixcomplement}
For spectral classifications that contain additional information beyond the spectral type and subtype, such as luminosity class and spectral peculiarities, we provide complementary data in the catalogue. This information is included only for targets that were not discarded during the encoding process described in Section~\ref{sec:cod_spec_types}. Each luminosity class, from hypergiants to subdwarfs, was assigned an integer value from 0 to 5 (as hypergiants and supergiants were grouped with the value of 0). For intermediate classes (e.g. II-III, II/III) the mean value, rounded to one decimal was adopted (for instance 2.5 was adopted for III/IV). In addition to the tuple, reported spectral peculiarity information was propagated by means of five, not mutually exclusive, flags (\textit{uncertain, miscellaneous, line-broadening, emission,} and \textit{nitrogen/carbon} enriched, see Table \ref{table:peculiarities})).

\begin{table}[ht]
  \begin{center}
    \caption{Categories of possible peculiarities:}
    \label{table:peculiarities}
    \begin{tabular}{ll}
    \hline
    Category & Peculiarity \\
    \hline
    \hline
    Uncertainty & :, ? \\
     \hline
     Miscellaneous & comp, k, p, pec, !, var, v, w, wl, wk \\
     \hline
     Line-broadening & (n), n, nn, nnn, [n], s, sh, shell \\
     \hline
     Emission lines & e, em, (e), [e], eq, neb, ((f)), (f), f, ((f*)), (f*), f*, c, f?p, (n)fp, nfp, z, q \\
     \hline
     Carbon/Nitrogen enhanced & N, C \\
    \hline
    \end{tabular}
  \end{center}
\end{table}

\section{Effects on the spectral type threshold in the BRF}\label{apendixB}
\begin{figure}[htbp!]
\centering
\includegraphics[scale=0.44]{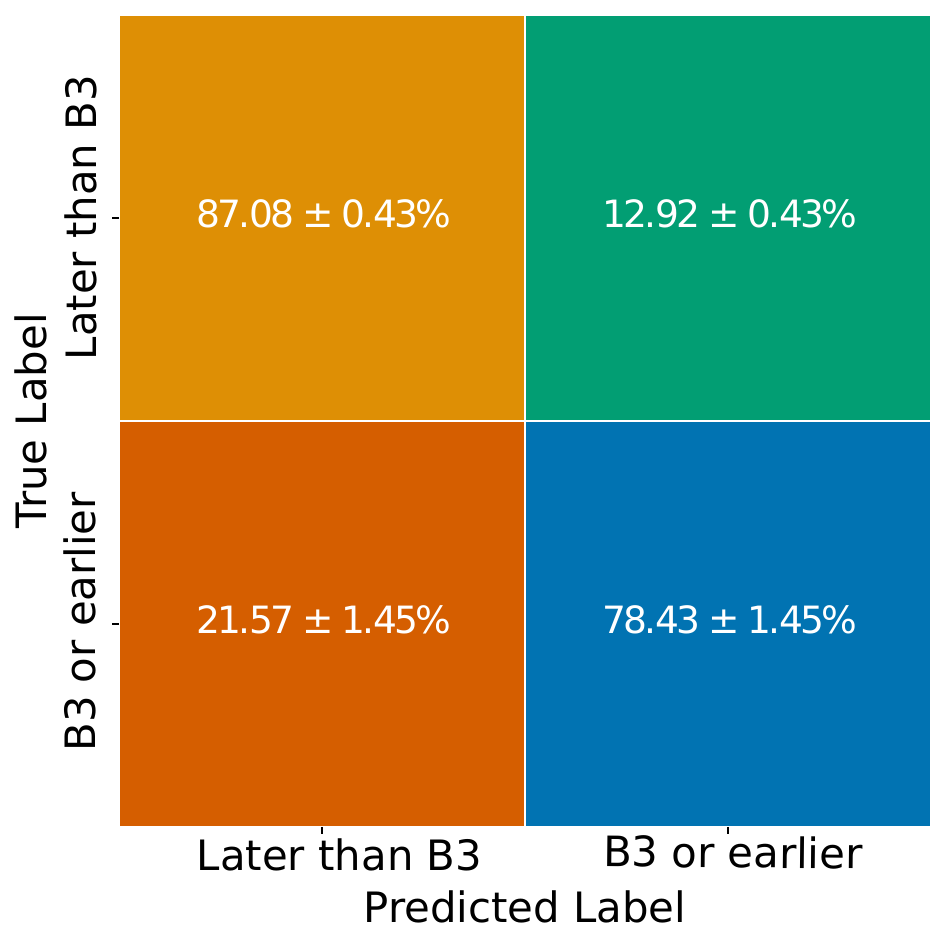}
\caption{Confusion matrix for stars earlier than B3 and later than B3. The confusion matrix was obtained by generating 20 distinct training and testing sets, and training independent models on each. After training, the models were applied to their respective testing sets. We report the mean and standard deviation across all testing sets.}
\label{fig:CM_colored_B3}
\end{figure}

\begin{figure}[htpb]
\centering
\includegraphics[scale=0.4]{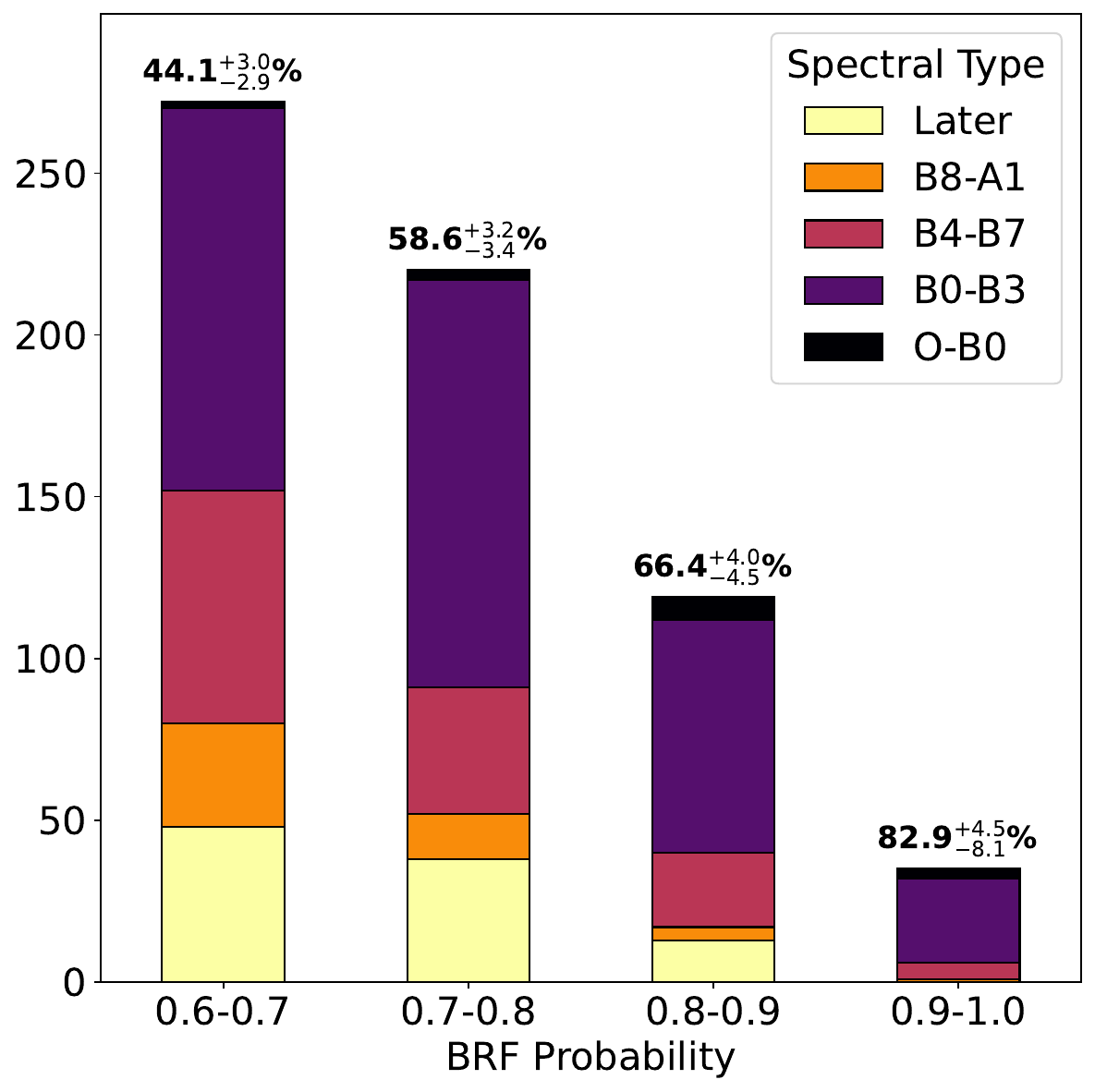}
\caption{Number of stars classified in different spectral type groups, divided by the probability given by the BRF using B3 tresholds candidates. The top of each bar shows the detection rate of massive stars.}
\label{fig:detection_rate_B3}
\end{figure}

The limit of massive stars can be different depending on the work and on the luminosity class, as we mention in Section \ref{sec:simplabel}. During the body of the work ``B2" was used as a threshold. However, in this appendix we show the impact that a slight change in threshold to ``B3" would have in our results. To this effect, we trained a new BRF following the same procedure described before but using B3 as the boundary between the two classes. The confusion matrix of this new model is shown in Fig. \ref{fig:CM_colored_B3}. In addition, we also performed the same flow described in the text for external validation with LAMOST spectroscopy. The final results of this external validation are shown in Fig.~\ref{fig:detection_rate_B3}.

Even though the difference in performance of the algorithm is not significant when comparing the two confusion matrices (Figs \ref{fig:CM_colored} and \ref{fig:CM_colored_B3}), the external validation procedure seems to indicate an underestimation of the contamination in the 0.6-0.7 probability bin for the case where B3 was used as a threshold.  Therefore, thresholding with "B2", provides more robust results.

\newpage
\section{}

\begin{table}[ht]
  \begin{center}
    \caption{Minimum and maximum values for different features.}
    \label{table:min_max_features}
    \begin{tabular}{lrr}
    \hline
    Feature & Min value & Max value \\
    \hline
    \hline
    $J$ & 5.531 & 14.69 \\
    $H$ & 5.502 & 14.69 \\
    $K_{s}$ & 5.462 & 14.70 \\
    $G$ & 5.665 & 14.97 \\
    $G_{BP}$ & 5.641 & 15.19 \\
    $G_{RP}$ & 5.611 & 14.73 \\
    $J$-$H$ & -0.125 & 0.433 \\
    $J$-$K_{s}$ & -0.154 & 0.661 \\
    $J$-$G$ & -2.426 & 0.3933 \\
    $J$-$G_{BP}$ & -3.246 & 0.4882 \\
    $J$-$G_{RP}$ & -1.434 & 0.2395 \\
    $H$-$K_{s}$ & -0.077 & 0.261 \\
    $H$-$G$ & -2.853 & 0.4913 \\
    $H$-$G_{BP}$ & -3.668 & 0.5923 \\
    $H$-$G_{RP}$ & -1.856 & 0.3399 \\
    $K_{s}$-$G$ & -3.086 & 0.5268 \\
    $K_{s}$-$G_{BP}$ & -3.892 & 0.6251 \\
    $K_{s}$-$G_{RP}$ & -2.087 & 0.3714 \\
    $G$-$G_{BP}$ & -0.8725 & 0.2467 \\
    $G$-$G_{RP}$ & -0.1592 & 1.046 \\
    $G_{BP}$-$G_{RP}$ & -0.2591 & 1.829 \\
    $\sigma_{\overline{\omega}}$ & 0.01088 & 0.4444 \\
    $\overline{\omega}/\sigma_{\overline{\omega}}$ & -2.518 & 161.8 \\
    $\overline{\omega}$ & -0.1438 & 5.356 \\
    $\mu_{\alpha}$ & -20.03 & 12.75 \\
    $\mu_{\delta}$ & -17.82 & 11.18 \\
    $\sigma_{\mu_{\delta}}$ & 0.01095 & 0.4344 \\
    $\sigma_{\mu_{\alpha}}$ & 0.00973 & 0.473 \\
    $ASM$ & 0.01693 & 0.7232 \\
    $AEN$ & 0.000 & 3.907 \\
    $RUWE$ & 0.6365 & 20.98 \\
    $AGOFAL$ & -9.306 & 265.8 \\
    $IPDGOFHA$ & 0.00265 & 0.1691 \\
    $\sigma_{AEN}$ & 0.000 & 1.564e+04 \\
    \hline
    \end{tabular}
  \end{center}
  \tablefoot{Minimum and maximum values for each feature used in the analysis.}
\end{table}

\begin{figure}
  \begin{center}
    \includegraphics[scale=0.7]{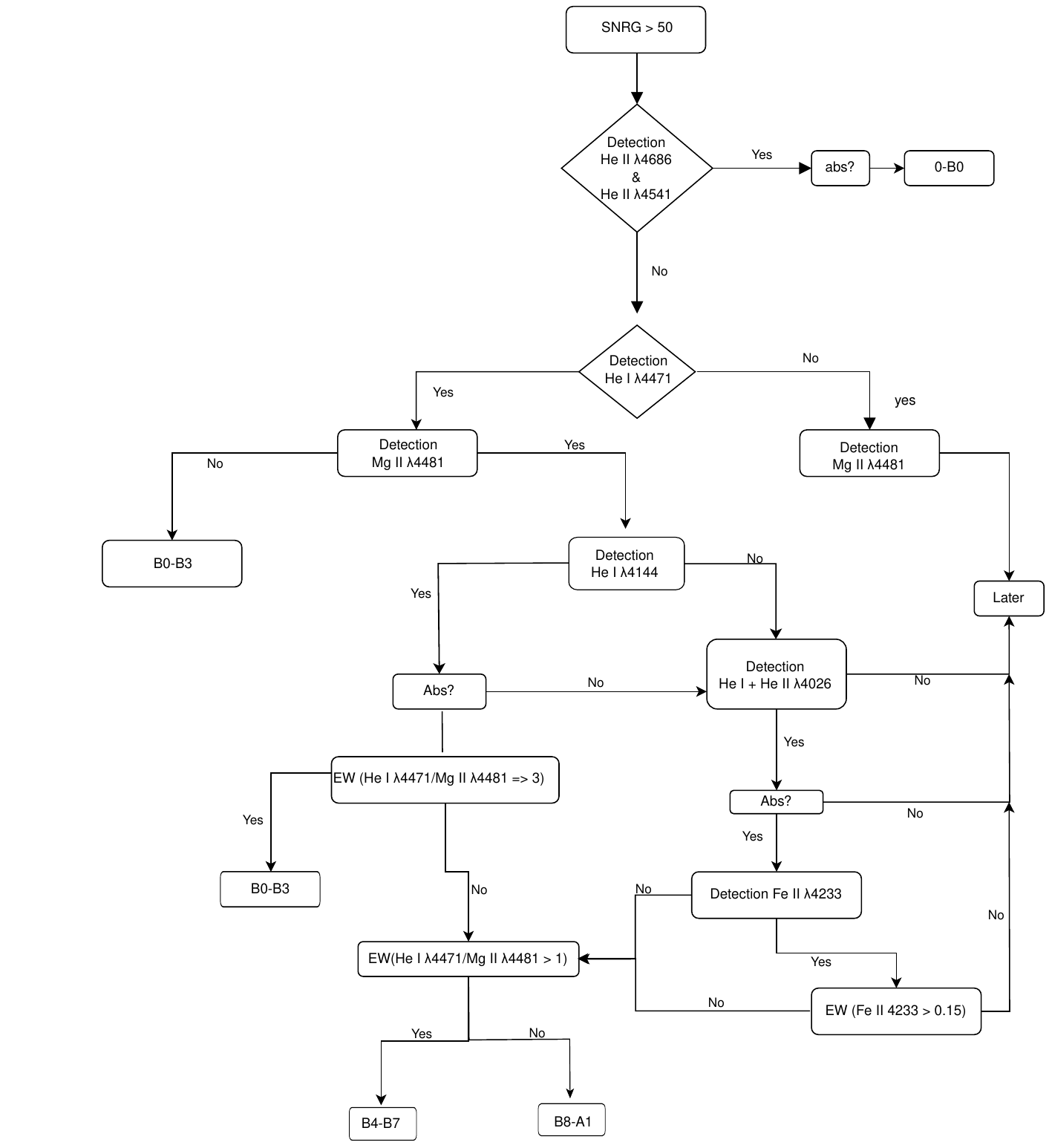}
  \end{center}

  \caption{Empirical decision tree based on the massive standard stars. The decision is based on the detection and equivalent width of the \ion{He}{i}, \ion{He}{ii}, \ion{Mg}{II}, and \ion{Fe}{II} lines at low resolution ($R \sim 1300$).}
  \label{fig:decision_clasificacion}
\end{figure}

\end{document}